\documentclass{JHEP3}

\usepackage{epsfig}
\usepackage{amsbsy}
\usepackage{varioref}
\usepackage{pifont}

\title{Next-to-leading Corrections to the Higgs Boson Transverse 
Momentum Spectrum in Gluon Fusion}

\author{Christopher J.~Glosser\\
 Department of Physics and Astronomy, University of Tennessee\\
Knoxville, TN 37996, USA\\
	E-mail: \email{cglosser@utk.edu}}

\author{Carl R.~Schmidt\\
Department of Physics and Astronomy, Michigan State University\\
East Lansing, MI 48824, USA\\
	E-mail: \email{schmidt@pa.msu.edu}}

\abstract{
We present a fully analytic calculation of the Higgs
boson transverse momentum and rapidity distributions, 
for nonzero Higgs $p_\perp$, at next-to-leading order
in the infinite-top-mass approximation.
We separate the cross section into a part that contains the
dominant soft, virtual, collinear, and small-$p_\perp$-enhanced
contributions, and the remainder, which is organized by the
contributions due to different parton helicities.  
We use this cross section to investigate analytically 
the small-$p_\perp$ limit and compare with the expectation 
from the resummation of large logarithms of the type $\ln{m_H/p_\perp}$.  
We also compute numerically the cross section at moderate
$p_\perp$ where a fixed-order calculation is reliable.
We find a $K$-factor that varies from $\approx1.6-1.8$,
and a reduction in the scale dependence, as compared to
leading order.  Our analysis suggests that the contribution of
current parton distributions to the total
uncertainty on this cross section at the LHC is probably less
than that due to uncalculated higher orders.
}

\keywords{QCD, Higgs Physics, Hadronic Colliders}
\preprint{{~MSUHEP--20915}\\
{~UTHEP--02--0902}\\
{~hep-ph/0209248}}

\begin{document}


\section{Introduction}

Our current understanding of particle physics depends crucially on the breaking of the
electroweak symmetry to give masses to the $W^\pm$ and $Z$ bosons, as well
as to all of the matter particles.  Yet, despite decades of extremely precise and 
successful predictions, the exact mechanism by which this symmetry breaking 
occurs is not known.  The simplest model involves a single weak $SU(2)$ scalar 
doublet which has a nonzero vacuum expectation value.  After rewriting in terms of the
physical states, this ``standard model''
 leaves behind a single neutral scalar, the Higgs boson, as its signature~\cite{Higgs}.  
Furthermore, models with extended Higgs sectors, including the minimal 
supersymmetric standard model (MSSM), often have a relatively light scalar with properties
similar to the standard model Higgs boson.  Therefore, the Higgs boson is at the top of any list of new particles to be found.

The direct search for the standard model Higgs boson in the $e^+e^-\rightarrow HZ$
channel at LEP2 has put a lower bound on its mass of 114.1 GeV~\cite{LEP2Limit}.  
Moreover, there are hints of a signal in the data just above this bound.
Meanwhile, precision electroweak measurements give an upper limit of
$m_H\lesssim 196-230$ GeV at the 95\% confidence level~\cite{HiggsRadCorr}.  Therefore, 
if a standard
model Higgs boson exists, its allowed mass range is not large.   Run II of the 
Tevatron can exclude a standard model Higgs boson over much of this mass
range, up to about 180 GeV, assuming 15 fb$^{-1}$ per experiment.
However, a definitive 5$\sigma$ discovery is difficult to obtain at this luminosity~\cite{RunIIExpectations} 
for a mass much beyond the LEP2 limit.
The Large Hadron Collider (LHC) at CERN will be needed to certify a Higgs discovery,
and to pin down its mass and couplings.

The dominant production mechanism for the Higgs boson at the LHC is the gluon-gluon fusion
process.  This process occurs at leading order (LO), ${\cal O}(\alpha_s^2)$, through a heavy quark
loop.
As is typical in Quantum Chromodynamic (QCD) processes initiated by gluons, the
radiative corrections are quite large.  
The next-to-leading order (NLO) 
corrections have been computed, including the full top-mass 
dependence~\cite{zerwas}, and indeed, 
the $K$-factor is $\approx 1.4-2.2$, depending on the Higgs mass
and the scale choice.  The computation with full $m_t$-dependence
 requires the calculation of two-loop diagrams and is quite complex. 
 Luckily, it simplifies greatly in the limit 
of large top quark mass ($m_t\rightarrow\infty$).  In this limit, one can integrate out the
top quark loop, leaving behind an effective gauge-invariant $Hgg$ 
vertex~\cite{HggVertex}.  
The Higgs boson production cross section has been calculated
at NLO in this limit in Refs.~\cite{Dawson}.
It gives an excellent approximation to the full $m_t$-dependent NLO cross
section for $m_H\lesssim 2m_t$.  Furthermore, the $K$-factor calculated in the
effective theory gives a good approximation to the full $m_t$-dependent
$K$-factor, even for larger Higgs masses.  Attempts have been made to
estimate the NNLO corrections using soft-gluon approximations~\cite{CdFG}
and resummations~\cite{Kramer}.  Recently the full NNLO cross section in the
large-$m_t$ limit has been computed~\cite{HK}.   Although the $K$ factor is
larger still at NNLO, the increase is not as severe as the NLO
enhancement, and the perturbation series seems to be well-behaved.

In addition to increasing the total cross section, the QCD radiation can have
a large effect on the kinematic distributions of the Higgs boson.  Most notably,
the transverse momentum, $p_\perp$, of the Higgs boson is exactly zero at LO, but 
is typically nonzero at higher orders, due to additional radiated partons.  
In fact, this additional QCD radiation has led some to consider searching for
the Higgs boson in association with a tagged jet at the LHC~\cite{Dubinin}.
Regardless of whether additional jets are tagged in the Higgs events, it is useful to understand
the transverse momentum distributions of both the Higgs signal and the 
background.
The transverse momentum spectrum of the Higgs boson has been calculated at
LO\footnote{Note that the LO contribution to the $p_\perp$ spectrum at
non-zero $p_\perp$ is actually down by $\alpha_s$ compared to the LO cross section;
it contributes to the cross section at NLO.}, both with the
full top quark mass dependence and in the large-$m_t$ limit~\cite{EHSvdB}.  
 It was seen that the large-$m_t$ limit is a good approximation
to this distribution if $m_H\lesssim 2m_t$ and $p_\perp\lesssim m_t$.  Recently, it was 
shown~\cite{DelDuca}
that these conditions are also sufficient to use the large-$m_t$ limit
for Higgs $+ 2$ jet production;  {\it i.e,} the transverse momenta of the Higgs boson
and the jets must be less than $\approx m_t$, but other invariants such as the total partonic
center-of-mass energy can still be large.

In this work we calculate the NLO corrections in the large-$m_t$ limit
 to the Higgs boson $p_\perp$ and rapidity spectrum, 
$d\sigma/dp_\perp^2/dy_H$, at the LHC.
Initial results of our calculation, which included the
purely gluonic contributions, were reported in Ref.~\cite{chris}.  
The NLO corrections in this limit have
been calculated previously in Ref.~\cite{dFGK}, using a Monte Carlo integration
to do the phase space integrations, after cancellation of the infrared 
singularities. More recently, a full analytic calculation of the $p_\perp$ 
spectrum was presented in Ref.~\cite{RSvN}.  Our calculation is 
similar to that in~\cite{RSvN} in that it is completely analytic.  
However, by treating the contributions of different helicity configurations
separately, we are able to report the full formulae for the differential
cross section in this paper,
in a relatively tractable form.  Another difference
is that we have used the $+$function technique to deal with soft 
singularities, whereas an artificial parameter was introduced in \cite{RSvN}
to separate the soft and hard radiative contributions.  Our technique
allows more of the universal structure of the NLO corrections to be 
apparent.  
Finally, since much of the total Higgs cross section occurs at not-too-large
$p_\perp$, we investigate the low $p_\perp$ limit of our result.

At small Higgs transverse momenta the perturbation series 
for the $p_\perp$ spectrum becomes unstable,
containing terms like $(\alpha_s^2/p_\perp^2)\alpha_s^n\ln^m(m_H^2/p_\perp^2)$,
with the leading logarithm occurring for $m=2n-1$.  This logarithmic series 
has been resummed, using the techniques of Collins, Soper, and 
Sterman~\cite{CSS}, at various levels of approximation~\cite{Hresum}.  
The NLO differential cross section at small $p_\perp$ contains
the fixed order terms in the logarithmic expansion corresponding to 
$n=2,\ m=3,2,1,0$.  In particular, the so-called $B^{(2)}$ coefficient in 
the logarithmic series occurs in our calculation at $n=2,\ m=0$.  This 
coefficient has recently been derived~\cite{dFG}
using the universality of the real emission contributions, combined with 
knowledge of the virtual correction amplitudes in the soft and collinear 
limits.  Using our analytic expressions for the NLO $p_\perp$ spectrum, 
we have verified this coefficient by direct calculation.  The 
analytic comparison of our cross section in the small-$p_\perp$ limit
with that expected from the resummation formulae is a very stringent
check on our results.

The remainder of this paper is organized as follows.
In section~\ref{sec:LO} we set up the calculation by defining some 
general formulae and giving the Born level expressions for the 
differential $p_\perp$ and rapidity spectrum.  In section~\ref{sec:NLO} 
we obtain the ${\cal O}(\alpha_s)$ corrections to this
 by combining the virtual one-loop (in the effective large-$m_t$ theory) 
amplitudes with the tree-level real radiative corrections.  Although
both of these contributions are infrared divergent, these divergences 
cancel after they are added together, using $\overline{\rm MS}$ parton 
density functions, defined at NLO.  In the main text we give formulae for 
the largest contribution to the distribution,
which contains all terms having singular behavior as one of the real 
QCD partons becomes soft or collinear, as well as most of the terms that 
are leading at small-$p_\perp$ (all but the $m=0$ terms from the previous
paragraph).  For lack of a better word, we label these contributions the
``singular'' contributions.  The remaining ``nonsingular'' contributions 
are given in appendix~\ref{app:nsrc}.   In section~\ref{sec:numbers} we 
give some numerical results and analysis obtained from our calculation, 
showing some representative distributions at the LHC.  
We also comment on the numerical comparison of our results to previous
calculations.  In 
section~\ref{sec:smallpt} we consider the small-$p_\perp$ limit 
of our result, and compute directly the $B^{(2)}$ coefficient.  Finally, 
in section~\ref{sec:conclusions} we give our conclusions.


\section{Higgs $p_\perp$ Spectrum: General Formulae and Leading Order expressions}
\label{sec:LO}

In the large-$m_t$ limit the top quark can be removed from the full theory, leaving a residual 
Higgs-gluon coupling term in the lagrangian of the effective theory:
\begin{equation}
{\cal L}_{\rm eff}\ =\ -{1\over4}\biggl[1-{\alpha_{s}\over3\pi}{H\over v}
\left(1+{\alpha_{s}\over4\pi}\Delta\right)\biggr]\,G^a_{\mu\nu}G^{a\mu\nu}\ .\end{equation}
The finite ${\cal O}(\alpha_{s})$ correction to this effective $Hgg$ 
operator, which
is necessary to the order we are working,
has been calculated \cite{Dawson} to be
\begin{equation}
\Delta\ =\ 5N_c-3C_F\ =\ {11}\ ,
	\label{opcorr}
\end{equation}
where $N_c=3$ and $C_F=(N_c^2-1)/(2N_c)=4/3$.
The ${\cal O}(\alpha_{s}^2)$ correction to the effective operator
has also been calculated~\cite{Chetyrkin:1997iv,Kramer}.

We consider the inclusive scattering of two hadrons $h_1$ and $h_2$ into a Higgs boson, 
$h_1h_2\rightarrow H+X$.  The differential cross section for this process, via the gluon-fusion 
production mechanism in perturbative QCD, for a Higgs boson of  
transverse momentum $p_\perp$ and rapidity $y_H$, can be written
\begin{equation}
	{d\sigma\over dp_{\perp}^{2}dy_{H}}
	\ =\  \sum_{i,j}	\int_{0}^{1}dx_{a}\,dx_{b}\,f_{i/h_1}(x_a,\mu_F)
	f_{j/h_2}(x_b,\mu_F){d\hat\sigma_{ij}\over dp_{\perp}^{2}dy_{H}}\ ,
\label{eq:master}
\end{equation}
where $i$ and $j$ label the massless partons which scatter from the hadrons $h_1$ and $h_2$,
respectively ($i=g,q_f,\bar q_f$, with $n_f$ flavors of light quarks) in the partonic subprocess $ij\rightarrow H+X$.   The parton densities
$f_{i/h}(x,\mu_F)$ are defined in the $\overline{\rm MS}$ factorization scheme at scale
$\mu_F$.  
The partonic subprocess cross section can be expanded as a power series in the strong coupling, $\alpha_s(\mu_R)$, as
\begin{equation}
	{d\hat\sigma_{ij}\over dp_{\perp}^{2}dy_{H}}\ =\ {\sigma_0\over\hat s}\left[
 {\alpha_s(\mu_R)\over2\pi}G^{(1)}_{ij}\,
+\,\left({\alpha_s(\mu_R)\over2\pi}\right)^2G^{(2)}_{ij}\,\ +\ \dots\right]\ ,
\label{eq:series}
\end{equation}
with
\begin{equation}
	\sigma_{0}\ =\ {\pi\over64}
	\left({\alpha_{s}(\mu_R)\over3\pi v}\right)^{2}\ ,
	\label{eq:sigmanought}
\end{equation}
where the Higgs vacuum expectation value $v$ is related to the Fermi constant,
$G_F=1.16639\times10^{-5}$ GeV$^{-2}$, by $v^{-2}=\sqrt{2}G_F$.
The quantities $G_{ij}^{(k)}$ are functions of the renormalization scale $\mu_R$,
the factorization scale $\mu_F$, the Higgs mass $m_H$, and the partonic Mandelstam
invariants, defined by
\begin{eqnarray}
	\hat s & = & (p_{a}+p_{b})^{2}\ =\ (p_{H}+Q)^{2}
	\ =\ sx_{a}x_{b}
	\label{eq:hats}  \nonumber\\
	\hat t & = & (p_{a}-Q)^{2}\ =\ (p_{b}-p_{H})^{2}
	\ =\ m_{H}^{2}-\sqrt{s}x_{b}m_{\perp}e^{y_{H}}
	\label{eq:hatt}  \\
	\hat u & = & (p_{b}-Q)^{2}\ =\ (p_{a}-p_{H})^{2}
	\ =\ m_{H}^{2}-\sqrt{s}x_{a}m_{\perp}e^{-y_{H}}\ ,
	\label{eq:hatu}\nonumber
\end{eqnarray}
where $p_a=x_ap_1$ and $p_b=x_bp_2$ are the initial-state parton momenta, $Q$ is the 
momentum of the final-state partons which balance the Higgs boson, and 
$m_{\perp}^{2}=m_{H}^{2}+p_{\perp}^{2}$.  These invariants satisfy 
\begin{equation}
	Q^{2}\ =\ \hat s+\hat t+\hat u-m_{H}^{2}\ .
	\label{eq:virtuality}
\end{equation}

At leading order the contributions of the different subprocesses (at nonzero $p_\perp$)
are given by 
\begin{equation}
G^{(1)}_{ij}\ =\ g_{ij}\,\delta(Q^2)
\end{equation}
 with
\begin{eqnarray}
	g_{gg} &=&  N_c\,\left({m_{H}^{8}
	+{\hat s}^{4}+{\hat t}^{4}+{\hat u}^{4}\over \hat u\hat t\hat s}\right)\nonumber\\
	g_{gq}
	&=& C_F\,\left({{\hat t}^{2}+\hat s^2\over-\hat u}\right)
	\label{eq:loforms}\\
	g_{q\bar{q}}
	&=& 2\,C_F^2\,\left({{\hat t}^{2}+\hat u^2\over\hat s}\right)\ ,\nonumber
\end{eqnarray}
 and $g_{qg}$ obtained from $g_{gq}$ via $\hat t\leftrightarrow\hat u$.  Note that the 
quark and antiquark in $g_{q\bar q}$ must be of the same flavor at leading order.


\section{${\cal O}(\alpha_s)$ Corrections}
\label{sec:NLO}

The ${\cal O}(\alpha_s)$ corrections come from three sources: the virtual corrections (V) to
Higgs-plus-one-parton production arising from the interference between
the Born and the one-loop amplitudes; the real corrections (R) from
Higgs-plus-two-parton production at the Born-level; and the Altarelli-Parisi corrections (AP)
arising from the definition of the $\overline{\rm MS}$ parton densities at NLO.

Although the total ${\cal O}(\alpha_s)$ correction to the inclusive Higgs production cross section 
is finite and well-defined in four dimensions, the individual V, R, and AP contributions are not and
must be calculated using some regularization procedure.  As is standard in perturbative QCD
we use conventional dimensional regularization (CDR) with dimension $d=4-2\epsilon$, so 
that the
collinear and soft singularities appear as $\epsilon$-poles in the separate contributions.  
The total ${\cal O}(\alpha_s)$ correction is 
\begin{equation}
G^{(2)}_{ij} \ =\ \lim_{\epsilon\rightarrow0}\Bigl(G^{(2{\rm V})}_{ij}(\epsilon)\ +\ G^{(2{\rm R})}_{ij}(\epsilon)\ +\ 
G^{(2{\rm AP})}_{ij}(\epsilon)\Bigr)\ .
\end{equation}

\subsection{Virtual Contributions}
\label{sec:vc}

The virtual one-loop expressions in the effective large-$m_t$ theory  were calculated in Ref.~\cite{Schmidt:1997wr}.  
Although they were calculated for helicity amplitudes in the t'Hooft-Veltman (HV)~\cite{HV}
and the 
four-dimensional-helicity schemes~\cite{fourDH}, they can be utilized in CDR by using the 
results of 
Catani, Seymour and Tr\'ocs\'anyi \cite{cst}. In that paper it was shown that the singularity
structure at NLO in the HV scheme and the CDR scheme are identical, when properly
defined, because the number of internal gluon helicities are taken to be the same; {\it i.e.,}
in $d=4-2\epsilon$ dimensions.
This implies that we can reconstruct the CDR virtual contributions by first constructing the virtual
corrections to the cross section in the HV scheme, using the amplitudes from Ref.~\cite{Schmidt:1997wr} with $\delta_R=1$, and then replacing the 4-dimensional
Born-level expressions $g_{ij}$ (\ref{eq:loforms}), which multiply all virtual singularities in the
HV scheme, by the following $d$-dimensional generalizations:
\begin{eqnarray}
	g_{gg}(\epsilon) &=&  {a_\epsilon\over(1-\epsilon)^2}\,N_c\,\left({(1-\epsilon)(m_{H}^{8}
	+{\hat s}^{4}+{\hat t}^{4}+{\hat u}^{4})-4\epsilon\, m_H^2\hat u\hat t\hat s\over \hat u\hat t\hat s}\right)\nonumber\\
	g_{gq}(\epsilon)
	&=& {a_\epsilon\over1-\epsilon}\,C_F\,\left({{\hat t}^{2}+\hat s^2-\epsilon\,(\hat t+\hat s)^2\over-\hat u}\right)
	\label{eq:loformsii}\\
	g_{q\bar{q}}(\epsilon)
	&=& a_\epsilon\,2\,C_F^2\,\left({{\hat t}^{2}+\hat u^2-\epsilon\,(\hat t+\hat u)^2\over\hat s}\right)\ ,\nonumber
\end{eqnarray}
where
\begin{equation}
a_\epsilon\ =\  {1\over\Gamma(1-\epsilon)}\left({4\pi\mu_R^2\over p_\perp^2}\right)^\epsilon\ .
\end{equation}
This accounts for the correct number of helicities of the external particles in the CDR scheme,
as well as the $d$-dimensional final-state phase space.

\subsection{Altarelli-Parisi Contributions}
\label{sec:ap}

The Altarelli-Parisi subtraction terms arise from the renormalization of the
parton density functions at NLO.  In the $\overline{\rm MS}$ scheme
they can be obtained from
\begin{eqnarray}
	G_{ij}^{(2{\rm AP})}(\epsilon)
	&=& {1\over\epsilon\Gamma(1-\epsilon)}\left({4\pi\mu_R^2\over\mu_F^2}\right)^{\epsilon}
	\sum_k
	\int_{0}^{1} 
	{dz\over z}\,\Biggl\{P_{ki}(z)g_{kj}(zx_a,x_b;\epsilon)\delta\Bigl(\hat t+(Q^2-\hat t)z\Bigr)
	\nonumber\\
	&&
	+ P_{kj}(z)g_{ik}(x_a,zx_b;\epsilon)\delta\Bigl(\hat u+(Q^2-\hat u)z\Bigr)
	\Biggr\}\ ,\label{eq:apterms}
\end{eqnarray}
where we have made explicit the functional dependence on $x_a$ and $x_b$
of the $g_{ij}(\epsilon)$, given in (\ref{eq:loformsii}).  
The splitting functions are
\begin{eqnarray}
	P_{gg}(z)&=& N_{c}\left[{1+z^{4}+(1-z)^{4}\over(1-z)_{+}z}\right]
	+{\beta_{0}}\delta(1-z)\nonumber\\
	P_{qq}(z)&=& C_F\left[{1+z^{2}\over(1-z)_{+}}+{3\over2}\delta(1-z)\right]
	\nonumber\\
	P_{gq}(z)&=& C_F\left[{1+(1-z)^{2}\over z}\right]
	\label{eq:splitgg}\\
	P_{qg}(z)&=& {1\over2}\left[z^{2}+(1-z)^{2}\right]\ ,	\nonumber
	\end{eqnarray}
where $\beta_0=(11N_c-2n_f)/6$ and $n_f$ is the number of light quarks. 
The $+$functions are defined by
\begin{equation}
	\int_{x}^{1}dz\,F(z)\left({g(z)}\right)_{+}
	\ =\ \int_{0}^{1}dz\,[F(z)\Theta(z-x)-F(1)]
	{g(z)}\ ,
	\label{eq:plus}
\end{equation}
where in eqs.~(\ref{eq:splitgg}) we have $g(z)=(1-z)^{-1}$.
The integrals over $z$ in (\ref{eq:apterms}) can be performed using the $\delta$-functions,
giving
\begin{equation}
	G_{ij}^{(2{\rm AP})}(\epsilon)
	\ =\  {1\over\epsilon\Gamma(1-\epsilon)}\left({4\pi\mu_R^2\over\mu_F^2}\right)^{\epsilon}
	\sum_k\Biggl[
{1\over -\hat t}\,P_{ki}(z_a)g_{kj,a}(z_a;\epsilon)+ {1\over -\hat u}P_{kj}(z_b)g_{ik,b}(z_b;\epsilon)	\Biggr]\ ,\label{eq:aptermsi}
\end{equation}
where
\begin{eqnarray}
	z_{a} & = & {-{\hat t}\over Q^{2}-{\hat t}}
	\nonumber  \\
	z_{b} & = & {-{\hat u}\over Q^{2}-{\hat u}}\ ,
	\label{eq:zb}
\end{eqnarray}
and we have defined $g_{ij,a}(z_a;\epsilon)\equiv g_{ij}(z_ax_a,x_b;\epsilon)$ and 
$g_{ij,b}(z_b;\epsilon)\equiv g_{ij}(x_a,z_bx_b;\epsilon)$.  
Note that $g_{ij,a}(z_a;\epsilon)$ can be obtained easily from the formulae of (\ref{eq:loformsii})
by the replacements $\hat s\rightarrow z_a\hat s,\,\hat t\rightarrow\hat t,\,\hat u\rightarrow  z_a\hat  sp_\perp^2/\hat t$. The expression $g_{ij,b}(z_b;\epsilon)$ can be obtained analogously.
For later use, we define
$g_{ij,a}(z_a)$ and $g_{ij,b}(z_b)$ as the $\epsilon\rightarrow0$ limit of these.

\subsection{Real Contributions}
\label{sec:rc}

The amplitudes
for the relevant Higgs-plus-two-parton-production processes 
at the Born level in the large-$m_t$ limit were calculated in Refs.~\cite{Dawson:1992au,Kauffman:1997ix}, using the helicity amplitude
method\footnote{There are several typographical errors in
the formulae for the squared matrix elements in the appendix of Ref.~\cite{Kauffman:1997ix}, which we list here.  The occurrences of 
$S_{34}$ in the denominator of eq.~(A12) should be replaced by $S_{24}$.
The expressions for $n_{13}$ and $n_{23}$ in eq.~(A16) should be replaced by
$n_{13}=-S_{24}n_{12}(1\leftrightarrow4)$ and $n_{23}=n_{13}(3\leftrightarrow4)$.  The interference term that occurs in the scattering of identical quark
pairs has the wrong sign on the right hand side of eq.~(A20).
}.  
Since we are using the CDR scheme at NLO, we must supplement the resulting 
4-dimensional squared-matrix elements, $\overline{|{\cal M}(d=4)|^2}$ 
(averaged over initial polarizations and colors and summed over
final ones), with the correct $\epsilon$-dependence
to obtain the squared-matrix elements in $d=4-2\epsilon$ dimensions, 
$\overline{|{\cal M}(d=4-2\epsilon)|^2}$; {\it i.e.,} we must add the quantity 
$\overline{|{\cal M}(d=4-2\epsilon)|^2}-\overline{|{\cal M}(d=4)|^2}$.
In practice this was done by investigating the collinear regions of the phase-space,
where correction terms, finite as $\epsilon\rightarrow0$, were obtained in accordance
with Ref.~\cite{cst}.  Most of these correction terms combine so that they
may be identified with ${\cal O}(\epsilon)$ parts of the splitting functions,
or universal terms multiplied by $\delta(Q^2)$.  The single remaining form
of correction term, which is not so easily identified, turns out to be 
crucial for obtaining the correct small-$p_\perp$ limit.  Thus, we are confident that our approach gives the correct CDR cross section at NLO.

The real contributions from initial-states $i$ and $j$ can be written
\begin{eqnarray}
G^{(2R)}_{ij}(\epsilon)&=& {1\over2\pi}\int_0^\pi d\phi\int_0^\pi d\theta\, (\sin\theta)^{1-2\epsilon}
(\sin\phi)^{-2\epsilon}\nonumber\\
&&\qquad\qquad\qquad\times
\left\{
\left({4\pi\mu_R^2\over p_\perp^2}\right)^\epsilon
\left({4\pi\mu_R^2\over Q^2}\right)^\epsilon
\left({\alpha_s\over3\pi v}\right)^{-2}
\left({\pi\alpha_s}\right)^{-2}
{\overline{|{\cal M}_{ij}|^2}\over\Gamma(1-2\epsilon)}
\right\}\,,\qquad
\end{eqnarray}
where $\theta$ and $\phi$ are the polar angles of the recoiling
partons in their center-of-mass frame, and in addition to
averages and sums over colors and polarizations, we also sum over
possible final-state flavors.  By
performing many partial fractionings on the squared-matrix elements,  
we can reduce the integrals to two types~\cite{Beenakker:1991ma}:
\begin{eqnarray}
\Omega^{(m,n)}_{kl}&=&{1\over2\pi}\int_0^\pi d\phi\int_0^\pi d\theta\, (\sin\theta)^{1-2\epsilon}
(\sin\phi)^{-2\epsilon}\, (S_{ak})^{-m}(S_{bl})^{-n}\label{inti}\\
\Omega^{(m,n)}_{akl}&=&{1\over2\pi}\int_0^\pi d\phi\int_0^\pi d\theta\,( \sin\theta)^{1-2\epsilon}
(\sin\phi)^{-2\epsilon}\, (S_{ak})^{-m}(S_{abl})^{-n}\ ,\label{intii}
\end{eqnarray}
where $S_{ak}=(p_a-p_k)^2$ and $S_{abk}=(p_a+p_b-p_k)^2$ with $a,b$ labeling 
the initial-state parton momenta and $k,l$ labeling the final-state parton 
momenta.  A general solution to
integrals of type (\ref{inti}) for arbitrary $\epsilon$ was given in Ref.~\cite{vanNeerven:1986xr}.
Solutions to the integrals of type (\ref{intii}) were given as a power series in $\epsilon$ in Ref.~\cite{Beenakker:1991ma}.
We have independently checked these integrals, and supplemented some of them to
retain terms of ${\cal O}(\epsilon)$ not given in~\cite{Beenakker:1991ma}.

The integral of the real-emission partonic cross section over the momentum fractions 
$x_a$ and $x_b$ in eq.~(\ref{eq:master}) has infrared singularities as $Q^{2}\rightarrow0$.
To handle these, we used the following identity:
\begin{equation}
	\left(Q^{2}\right)^{-1-n\epsilon} \ =\ 
	  - {1\over n\epsilon}
	\delta(Q^{2})\left({-\hat t}\right)^{-n\epsilon}+
	 {z_{a}\over-\hat t}
	 \left[{\left({-\hat t}/z_{a}\right)^{-n\epsilon}
	 \over(1-z_{a})_{+}}-
	 n\epsilon\left({\ln(1-z_{a})\over1-z_{a}}\right)_{+}+\cdots\right]
	 \nonumber \ ,
	\nonumber
\end{equation}
and the equivalent equation obtained by $a\leftrightarrow b$ and ${\hat t}
\leftrightarrow {\hat u}$. 

To organize the final expressions, we write the real contributions as a sum 
of ``singular'' and ``nonsingular'' terms:
\begin{equation}
G^{(2{\rm R})}_{ij}(\epsilon) \ =\ G^{(2{\rm R,s})}_{ij}(\epsilon)\ +\ G^{(2{\rm R,ns})}_{ij}\ ,
\end{equation}
where the ``nonsingular'' terms are finite as $\epsilon\rightarrow0$ and as $Q^2\rightarrow0$.  
We also choose to define the separation so that they have no naive singularities as $p_\perp
\rightarrow0$ as well, prior to the integration over $x_a$ and $x_b$ in Eq.~(\ref{eq:master});
this ensures that the ``nonsingular'' terms are not enhanced by any power of
$\ln(m_H^2/p_\perp^2)$ in the small-$p_\perp$ limit.  (They will have $1/p_\perp^2$ 
singularities after the integration over $x_a$ and $x_b$ , but without logarithmic enhancement.)  
Explicit formulae for the ``nonsingular'' terms
are given in appendix~\ref{app:nsrc}.  

\subsection{Total Corrections}
\label{sec:total}

After combining the three contributions, the $\epsilon$-poles cancel,
and we can safely set the 
dimensional regularization parameter $\epsilon$ to zero.  We can then write
\begin{equation}
G^{(2)}_{ij} \ =\ G^{(2{\rm s})}_{ij}\ +\ G^{(2{\rm R,ns})}_{ij}\ ,
\end{equation}
where the ``singular'' terms, $G^{(2{\rm s})}_{ij}$, contain the sum of the virtual, the Altarelli-Parisi, and the ``singular'' real contributions.  In this section we give formulae for the $G^{(2{\rm s})}_{ij}$, which contain
by far the dominant contribution to the cross section.   In the small $p_\perp$ limit
they contain all of the contributions enhanced by $(\alpha_s^2/p_\perp^2)\alpha_s^2\ln^m(m_H^2/p_\perp^2)$, with $m=3,2,1$.  
The ``nonsingular'' terms, $G^{(2{\rm R,ns})}_{ij}$, which we give in appendix~\ref{app:nsrc}, contain only subleading $m=0$ contributions in this limit . 

In the following equations we use the quantity $Q^2_\perp=Q^2+p_\perp^2$ 
and the definitions
\begin{eqnarray}
	p_{gg}(z)&=& (1-z)P_{gg}\ =\ N_{c}\left[{1+z^{4}+(1-z)^{4}\over z}\right]
	\nonumber\\
	p_{qq}(z)&=& (1-z)P_{qq}\ =\  C_F\left[{1+z^{2}}\right]\ ,
	\end{eqnarray}
and
\begin{eqnarray}
	C^\epsilon_{gg}(z)&=& 0
	\nonumber\\
	C^\epsilon_{qq}(z)&=& C_F\,(1-z)	\nonumber\\
	C^\epsilon_{gq}(z)&=& C_F\,z	\\
	C^\epsilon_{qg}(z)&=& z(1-z)\ .\nonumber
	\end{eqnarray}
Note that the $C^\epsilon_{ij}$ are minus the ${\cal O}(\epsilon)$ parts of the splitting functions.

For the gluon-gluon singular term we obtain
\begin{eqnarray}
G^{(2{\rm s})}_{gg} &=&
\delta(Q^2)\Biggl\{\left(\Delta +\delta+N_cU\right)g_{gg}
	\nonumber  \\
	&&+
	(N_{c}-n_{f}){N_c\over3}\left[(m_{H}^{4}/\hat s)+(m_{H}^{4}/\hat t)
+(m_{H}^{4}/\hat u)+m_{H}^{2}
\right]\Biggr\}+\Biggl\{\nonumber\\
&&\quad\left({1\over -\hat t}\right)\Biggl[
-P_{gg}(z_a)\ln{\mu_F^2z_a\over(-\hat t)}
+p_{gg}(z_a)\left({\ln1-z_a\over1-z_a}\right)_+\Biggr]g_{gg,a}(z_a)\nonumber\\
&&+\left({1\over -\hat t}\right)\Biggl[
-2n_f\,P_{qg}(z_a)\ln{\mu_F^2\over Q^2}+
2n_f\,C^\epsilon_{qg}(z_a)\Biggr]g_{qg,a}(z_a)\nonumber\\
&&+\left({z_a\over -\hat t}\right)\left(\left({\ln1-z_a\over1-z_a}\right)_+
-{\ln\left( Q_\perp^2z_a/(-\hat t)\right)\over(1-z_a)_+}\right)\nonumber\\
&&\qquad\times {N_c^2\over2}\left[{\left(
m_H^8+\hat s^4+Q^8+\hat u^4+\hat t^4\right)+z_az_b\left(m_H^8+\hat s^4+Q^8+(\hat u/z_b)^4+(\hat t/z_a)^4\right)\over\hat s\hat u\hat t}\right]\nonumber\\
&&-\left({z_a\over -\hat t}\right)\left({1\over1-z_a}\right)_+{\beta_0\over2}N_c
\left({m_H^8+\hat s^4+z_az_b\left((\hat u/z_b)^4+(\hat t/z_a)^4\right)
\over\hat s\hat u\hat t}\right)\nonumber\\
&&+\Bigl[(\hat t,a)\leftrightarrow(\hat u,b)\Bigr]\Biggr\}\nonumber\\
&&+N_c^2\left[
{\left(m_H^8+\hat s^4+Q^8+(\hat u/z_b)^4+(\hat t/z_a)^4\right)\left(Q^2+Q_\perp^2\right)\over\hat s^2Q^2Q_\perp^2}\right.\nonumber\\
&&\qquad\qquad\qquad\left.+{2m_H^4\left((m_H^2-\hat t)^4 +(m_H^2-\hat u)^4+\hat u^4+\hat t^4\right)
\over\hat s\hat u\hat t(m_H^2-\hat u)(m_H^2-\hat t)}\right]\,{1\over p_\perp^2}\ln {p_\perp^2\over Q_\perp^2}\ ,\label{eq:ggsing}
\end{eqnarray}
where 
\begin{equation}
\delta\ =\ {3\beta_0\over2}\,
\left(\ln{\mu_R^2\over-\hat t}
+\ln{\mu_R^2\over-\hat u}\right)
\,+\,\left({67\over18}N_c-{5\over9}n_f\right)\ ,
\end{equation}
and
\begin{eqnarray}
U&=& 
{1\over2}\ln^2{-\hat u\over-\hat t}
\,+\,{\pi^2\over3}\\ 
&&-\,\ln {{\hat s}\over m_{H}^{2}}
\ln{-{\hat t}\over m_{H}^{2}}
\,-\,\ln{{\hat s}\over m_{H}^{2}}
\ln {-{\hat u}\over m_{H}^{2}}
\, -\,\ln {-{\hat t}\over m_{H}^{2}}
\ln {-{\hat u}\over m_{H}^{2}}
  \nonumber\\
&&+\ln^{2}{m_{H}^{2}\over{\hat s}}
+\ln^{2}{m_{H}^{2}\over m_{H}^{2}-{\hat t}}
+\ln^{2}{m_{H}^{2}\over m_{H}^{2}-{\hat u}}\\
&&+\,2\,{\rm Li}_2\biggl({{\hat s}-m_{H}^{2}\over {\hat s}}\biggr)
 \,+\,2\,{\rm Li}_2\biggl({m_{H}^{2}\over m_{H}^{2}-{\hat t}}\biggr)
 \,+\,2\,{\rm Li}_2\biggl({m_{H}^{2}\over m_{H}^{2}-{\hat u}}\biggr)
\ ,\nonumber
 \label{Universaltot}
\end{eqnarray}
Here, ${\rm Li}_2$ is the dilogarithm function.

The gluon-quark singular term is
\begin{eqnarray}
G^{(2{\rm s})}_{gq} &=&
\delta(Q^2)\Biggl\{\left(\Delta +N_cV_{1}+C_FV_{2}+V_{3}\right)g_{gq}
	\nonumber  \\
	&&+
(N_{c}-C_F)C_F\left[{\hat s^2+\hat t^2+\hat u^2-\hat um_H^2\over-\hat u}\right]\Biggr\}
    \nonumber\\
 &&\quad	\left({1\over -\hat t}\right)\Biggl[-P_{gg}(z_a)\ln{\mu_F^2z_a\over(-\hat t)}+p_{gg}(z_a)\left({\ln1-z_a\over1-z_a}\right)_+
\Biggr]g_{g q,a}(z_a)\nonumber\\
&&+\left({1\over -\hat t}\right)\Biggl[-
P_{qg}(z_a)\ln{\mu_F^2\over Q^2}+C^\epsilon_{qg}(z_a)\Biggr]g_{q\bar q,a}(z_a)\nonumber\\
&&+\left({1\over -\hat u}\right)\Biggl[-P_{qq}(z_b)\ln{\mu_F^2z_b\over(-\hat u)}+p_{qq}(z_b)\left({\ln1-z_b\over1-z_b}\right)_+
+C^\epsilon_{qq}(z_b)\Biggr]g_{gq,b}(z_b)\nonumber\\
&&+\left({1\over -\hat u}\right)\Biggl[-P_{gq}(z_b)\ln{\mu_F^2\over Q^2}+C^\epsilon_{gq}(z_b)\Biggr]g_{gg,b}(z_b)\nonumber\\
&&+\left({z_a\over -\hat t}\right)\left(\left({\ln1-z_a\over1-z_a}\right)_+
-{\ln\left( Q_\perp^2z_a/(-\hat t)\right)\over(1-z_a)_+}\right)\nonumber\\
&&\qquad\times N_cC_F\left[{
-\hat s^3\hat t-\hat s\hat t^3+Q^6\hat t+Q^2\hat t^3
\over\hat s\hat u\hat t}\right.\nonumber\\
&&\qquad\qquad\qquad\left.+{z_az_b\left(
-\hat s^3(\hat t/z_a)-\hat s(\hat t/z_a)^3-Q^6(\hat u/z_b)-Q^2(\hat u/z_b)^3
\right)\over\hat s\hat u\hat t}\right]\nonumber\\
&&-\left({z_b\over -\hat u}\right)\left({1\over1-z_b}\right)_+
{3\over2}C_F^2\left({\hat t^2+\hat s^2\over-\hat u}\right)\nonumber\\
&&+N_cC_F\left[
{\left(
-\hat s^3(\hat t/z_a)-\hat s(\hat t/z_a)^3-Q^6(\hat u/z_b)-Q^2(\hat u/z_b)^3
\right)(Q^2+Q_\perp^2)\over\hat s^2Q^2
Q_\perp^2}\right.\nonumber\\
&&\qquad\qquad\left.-{2m_H^4\left((m_H^2-\hat t)^2+\hat t^2\right)
\over\hat s\hat u(m_H^2-\hat u)}\right]\,{1\over p_\perp^2}\ln {p_\perp^2\over Q_\perp^2} \label{eq:gqsing}
\end{eqnarray}
with
\begin{eqnarray}
{ V_{1}}&=&
{1\over2}\ln^2{-\hat u\over-\hat t}
+{1\over2}\ln^2{\hat s\over-\hat u}-{1\over2}\ln^2{\hat s\over-\hat t}\nonumber\\
&&+\ln {{\hat s}\over m_{H}^{2}}
\ln{-{\hat t}\over m_{H}^{2}}
-\,\ln {{\hat s}\over m_{H}^{2}}
\ln{-{\hat u}\over m_{H}^{2}}
\,-\,\ln{-{\hat t}\over m_{H}^{2}}
\ln{-{\hat u}\over m_{H}^{2}}
  \\
&&+\,2\,{\rm Li}_2\biggl({m_H^2\over m_H^2-{\hat u}}\biggr)
\,+\,\ln^{2}{m_{H}^{2}\over m_{H}^{2}-{\hat u}}
\,+\,{\pi^{2}}
 \label{Universal1it}\nonumber\\
 V_{2}&=&
\ln^{2}{m_{H}^{2}\over{\hat s}}
+\ln^{2}{m_{H}^{2}\over m_{H}^{2}-{\hat t}}\,-2\,\ln {{\hat s}\over m_{H}^{2}}
\ln{-{\hat t}\over m_{H}^{2}}\nonumber\\
 &&+\,2\,{\rm Li}_2\biggl({{\hat s}-m_{H}^{2}\over {\hat s}}\biggr)
 \,+\,2\,{\rm Li}_2\biggl({m_{H}^{2}\over m_{H}^{2}-{\hat t}}\biggr)
-\,{7\over2}\,-\,{2\pi^{2}\over3}\ , \label{Universal2it}\\
 V_{3}&=&{\beta_0}
\left(2\ln{\mu_R^2\over-\hat u}+\ln{\mu_R^2\over-\hat t}\right)
\,+\,\left({67\over9}N_c-{10\over9}n_f\right) \ .\label{Universal3it}
\end{eqnarray}

For the quark-antiquark terms we must now distinguish between the scattering of
identical or nonidentical flavors.
The quark-antiquark singular term (same flavor) is
\begin{eqnarray}
G^{(2{\rm s})}_{q_i\bar q_i} 
&=&\delta(Q^2)\Biggl\{\left(\Delta +N_cW_{1}+C_FW_{2}+W_{3}\right)g_{q\bar q}
	\nonumber  \\
	&&+
	(N_{c}-C_F)\,2C_F^2\left[{\hat t^2+\hat u^2+\hat s^2-\hat sm_H^2
\over\hat s}\right]\Biggr\}\,
+\,\Biggl\{\nonumber\\
 &&
\quad \left({1\over -\hat t}\right)\Biggl[-P_{qq}(z_a)\ln{\mu_F^2z_a\over(-\hat t)}+p_{qq}(z_a)
\left({\ln1-z_a\over1-z_a}\right)_++C^\epsilon_{qq}(z_a)
\Biggr]g_{q\bar q,a}(z_a)\nonumber\\
&&+\left({1\over -\hat t}\right)\Biggl[
-P_{gq}(z_a)\ln{\mu_F^2\over Q^2}+C^\epsilon_{gq}(z_a)\Biggr]g_{g q ,a}(z_a)\nonumber\\
&&+\left({z_a\over -\hat t}\right)\left(\left({\ln1-z_a\over1-z_a}\right)_+
-{\ln \left(Q_\perp^2z_a/(-\hat t)\right)\over(1-z_a)_+}\right)\nonumber\\
&&\qquad\times\, \left(2C_F-N_c\right)\,C_F^2\,\left[{
\hat t^2+\hat u^2+(\hat t/z_a)^2+(\hat u/z_b)^2
\over\hat s}\right]\nonumber\\
&&-\left({z_a\over -\hat t}\right)\left({1\over1-z_a}\right)_+
{\beta_0}C_F^2\left({\hat t^2+\hat u^2\over\hat s}\right)\nonumber\\
&&+\Bigl[(\hat t,a)\leftrightarrow(\hat u,b)\Bigr]\Biggr\}\nonumber\\
&&+{2C_F^2}\left[{ (\hat s-Q^2)^2+\left(\hat u+\hat t-2Q^2\right)^2
\over\hat s}
\right]\,{1\over p_\perp^2}\ln {p_\perp^2\over Q_\perp^2}\ ,\label{eq:qiaising}
\end{eqnarray}
with
\begin{eqnarray}
{ W_{1}}&=&
\ln {-{\hat u}\over m_{H}^{2}}
\ln{-{\hat t}\over m_{H}^{2}}
-\,\ln{{\hat s}\over m_{H}^{2}}
\ln{-{\hat u}\over m_{H}^{2}}
\,-\,\ln {{\hat s}\over m_{H}^{2}}
\ln {-{\hat t}\over m_{H}^{2}}
  \nonumber\\
&&+\,2\,{\rm Li}_2\biggl({{\hat s}-m_{H}^{2}\over {\hat s}}\biggr)
\,+\,\ln^{2}{m_{H}^{2}\over {\hat s}}
\,-\,{1\over2}\ln^{2}{-\hat u\over -{\hat t}}
\,-\,{5\pi^{2}\over3}
 \label{Universal1iit}\\
 W_{2}&=&
{3\over2}\left[\ln {\hat s\over-{\hat t}}+\ln {\hat s\over-{\hat u}}\right]
 +\ln^{2}{-\hat u\over -{\hat t}}
\,-2\,\ln {-{\hat u}\over m_{H}^{2}}
\ln{-{\hat t}\over m_{H}^{2}}
 \nonumber\\
&&+\ln^{2}{m_{H}^{2}\over m_{H}^{2}-{\hat u}}
+\ln^{2}{m_{H}^{2}\over m_{H}^{2}-{\hat t}}
\\
 &&+\,2\,{\rm Li}_2\biggl({m_{H}^{2}\over m_{H}^{2}-{\hat u}}\biggr)
\,+\,2\,{\rm Li}_2\biggl({m_{H}^{2}\over m_{H}^{2}-{\hat t}}\biggr)
-\,7\,+\,{2\pi^{2}}\ ,\nonumber\\
 \label{Universal2iit}
 W_{3}&=&{\beta_0\over2}
\left(4\ln{\mu_R^2\over\hat s}+\ln{\mu_R^2\over-\hat u}+\ln{\mu_R^2\over-\hat t}\right)
\,+\,\left({67\over6}N_c-{5\over3}n_f\right) \ .\label{Universal3iit}
\end{eqnarray}

Finally, the quark-antiquark singular term (different flavors) is
\begin{eqnarray}
G^{(2{\rm s})}_{q_i\bar q_j} 
 &=&\Biggl\{	\left({1\over -\hat t}\right)\Biggl[-
P_{gq}(z_a)\ln{\mu_F^2\over Q^2}+C^\epsilon_{gq}(z_a)\Biggr]g_{g q,a}(z_a)\nonumber\\
&&\quad
+\Bigl[(\hat t,a)\leftrightarrow(\hat u,b)\Bigr]\Biggr\}\nonumber\\
&&
+{2C_F^2}\left[{ (\hat s-Q^2)^2+\left(\hat u+\hat t-2Q^2\right)^2
\over\hat s}
\right]\,{1\over p_\perp^2}\ln {p_\perp^2\over Q_\perp^2}
\ .\label{eq:qiajsing}
\end{eqnarray}
The singular terms for the quark-quark scattering, both same flavor or different flavor 
quarks, are identical to (\ref{eq:qiajsing}).


\section{Numerical Results and Analysis}
\label{sec:numbers}

In this section we present some numerical results of our 
calculation.  A convenient choice of variables that we use 
for the integrals over $x_a$ and $x_b$ is given in 
appendix~\ref{app:phasespace}.
We choose to make all plots for proton-proton collisions at
the LHC center-of-mass
energy of $\sqrt{s}=14$ TeV, and for a Higgs mass of
$m_H=120$ GeV.  This value of the Higgs mass is not far above
the direct search limits found by LEP2.  Unless otherwise
indicated, we use as a standard choice of renormalization
and factorization scales, $\mu_R=\mu_F=m_\perp$.  For parton density
functions (PDFs) we use the CTEQ5M1 distributions for all NLO
plots and the CTEQ5L distributions for all LO plots~\cite{CTEQ},
with the corresponding values of $\alpha_s(m_Z)=0.118$ and 
$\alpha_s(m_Z)=0.127$, respectively (with $\alpha_s$ running
at corresponding order).  We discuss the consequences of this choice
and the error due to PDFs in general below.

Before presenting our results we must first understand
the range of Higgs $p_\perp$ for which they are applicable.
In Fig.~\ref{fig:locomparison} we plot a comparison of 
the $p_\perp$ spectrum for $y_H=0$ computed at LO,
with the exact $m_t$ dependence
(dashes), and in the $m_t=\infty$ effective theory (solid).
For $p_\perp\lesssim200$ GeV the effective theory underestimates
the cross section by about 5 percent, which, as we shall see, is
much less than the NLO corrections. Above 200 GeV, the
exact calculation and the effective theory calculation begin
to diverge, with the effective theory calculation overestimating
the cross section.  This behavior is similar for other
Higgs boson masses as long as $m_H\lesssim250$ GeV, where the
effective theory underestimates the cross section by about
25 percent in the low $p_\perp$ region.
Thus, we assume that our NLO calculation
will be good for 
$p_\perp\lesssim200$ GeV, and is applicable for the Higgs masses
that are favored by the electroweak data.  From here
on, all of our plots  will be calculated in the $m_t=\infty$ 
effective theory.

The effect of NLO corrections on the Higgs $p_\perp$ and
rapidity distributions have been studied numerically
before in Refs.~\cite{RSvN}
and \cite{dFGK}.  Since these two calculations appear to be
in agreement\footnote{Although the version of Ref.~\cite{RSvN}
published in Nucl.\ Phys.\ B {\bf 634}, 247 (2002)
claims some disagreement with 
Ref.~\cite{dFGK}, the most recently revised version of their preprint,
{\tt hep-ph/020114v4}, which appeared after the published version,
now agrees with~\cite{dFGK}.  The change was due to a wrong implementation
of the MRST PDFs in the earlier version of~\cite{RSvN}.}, 
we have only checked our calculation
directly against that of Ref.~\cite{RSvN}.  Using MRST99 PDFs, and
rescaling the effective $Hgg$ coupling to be consistent with their definition,
we have been able to reproduce Figs. 3 and 4 from~\cite{RSvN} with
excellent agreement.

\EPSFIGURE[p]{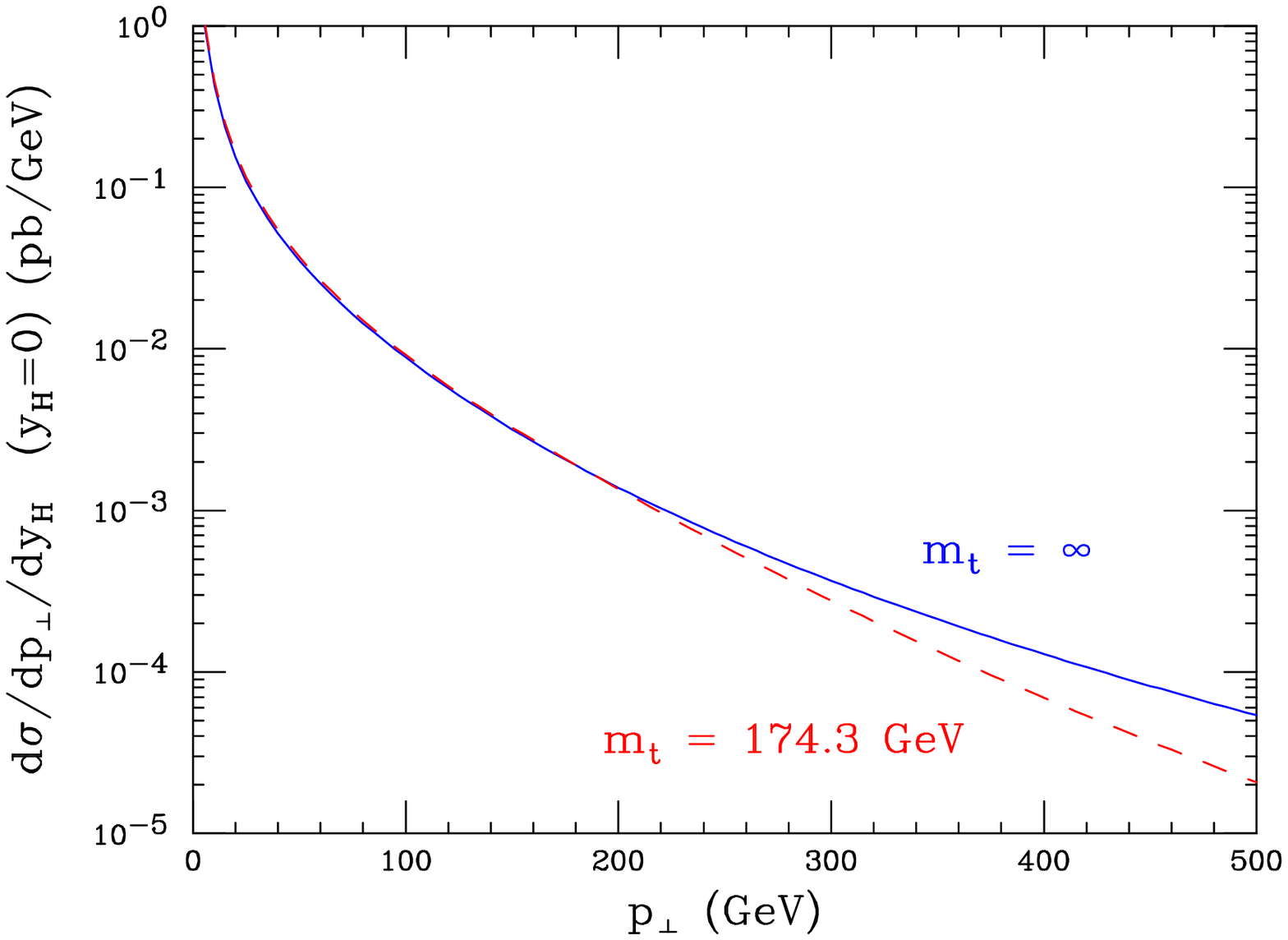,width=0.8\textwidth}
{The Higgs $p_\perp$ spectrum for $y_H=0$ 
calculated at LO, with the exact $m_t$ dependence
(dashes), and in the $m_t=\infty$ effective theory (solid).
\label{fig:locomparison}}

In Fig.~\ref{fig:ptY} we plot the Higgs $p_\perp$ spectrum
for three different values of the Higgs rapidity $y_H$.
These curves show the standard features of this distribution:
the fall off with higher $p_\perp$ and larger $y_H$, and the
steep plummet of the curve at very small $p_\perp\ll m_H$.
This last feature is due to the large logarithms $\ln(m_H/p_\perp)$,
which give a large negative contribution at NLO at small $p_\perp$.
Although the fixed order perturbation theory is unstable in this region,
below about $p_\perp=30$ GeV, we continue the plot here
because we are interested in the connection between the fixed-order
cross section and the resummation of the large logarithms.
This will be discussed in more detail in section~\ref{sec:smallpt}.

\EPSFIGURE[p]{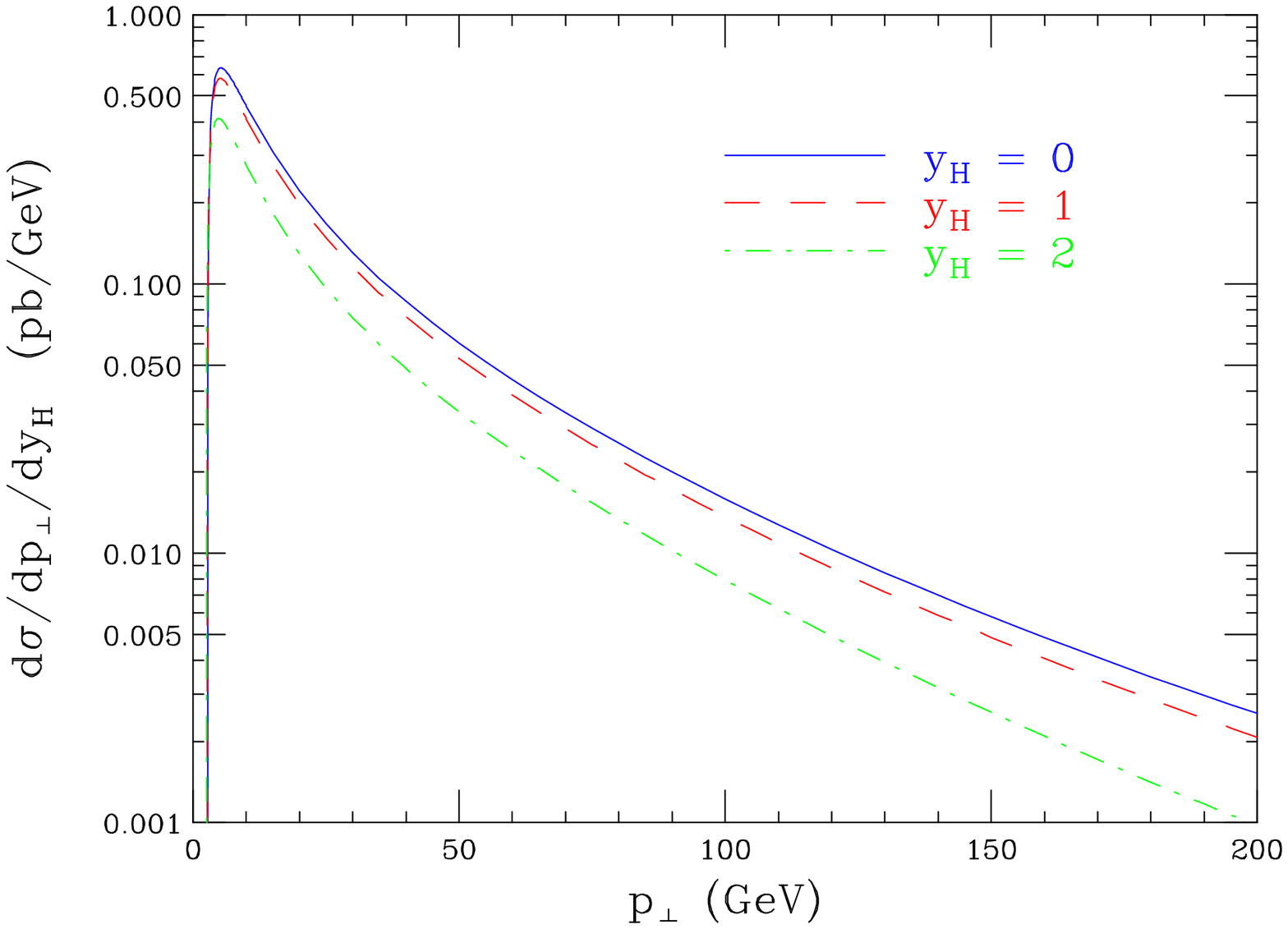, width=0.8\textwidth}
{The Higgs $p_\perp$ spectrum at NLO in the
$m_t=\infty$ effective theory for
rapidities $y_H=0$ (solid), $y_H=1$ (dashes), and 
$y_H=2$ (dotdash).
\label{fig:ptY}}

In Fig.~\ref{fig:ptsubcomponents} we plot the separate contribution
of the different initial-state partons to the $p_\perp$ spectrum
at $y_H=0$.  At small $p_\perp$ the cross section is strongly dominated
by the gluon-gluon component, whereas at large $p_\perp\approx 200$ GeV
the (anti)quark-gluon contribution makes up about 37\% of the total
cross section.  The (anti)quark-(anti)quark contribution, summed
over all flavors, is very small and negative over most of 
the range of this plot.  Its magnitude is less than one 
percent of the total for $p_\perp>30$ GeV, where the fixed order
perturbation theory is reliable.  

In Fig.~\ref{fig:ptbands} we plot a comparison of the LO
distribution with the NLO distribution for $y_H=0$, while varying
the renormalization and fragmentation scales together 
($\mu_R=\mu_F$) between $0.5m_\perp$ and $2m_\perp$.
This exhibits the two main features of the NLO cross section, 
relative to LO:  it is substantially larger, and the
uncertainty due to scale variation is smaller.
We illustrate these two features in more detail in 
Figs.~\ref{fig:ptKfactor} and \ref{fig:ptScale}.
In Fig.~\ref{fig:ptKfactor} we plot the $K$-factor, defined
as the ratio of the differential cross section calculated at
NLO and LO; that is,
\begin{equation}
K\ =\ {\left(d\sigma/dp_\perp^2/dy_H\right)_{\rm NLO}\over
\left(d\sigma/dp_\perp^2/dy_H\right)_{\rm LO}}\ .
\end{equation}
In this plot we have set the scales to $\mu_R=\mu_F=m_\perp$.
We also reiterate that the LO cross 
section in this ratio has been calculated using the LO
PDFs CTEQ5L.  For $p_\perp\gtrsim50$ GeV, the $K$-factor
is relatively constant, rising slowly from 1.7 to more than
1.8.  
This is comparable to the NLO $K$-factor for the total
cross section, for this value of the Higgs mass, $m_H=120$ GeV.
For $p_\perp\lesssim50$ GeV the $K$-factor begins to drop more
dramatically, less than 1.6 at $p_\perp=30$ GeV, where the
large logarithms start to become important.  Below this, the 
$K$-factor plummets, indicating the inapplicability of this
fixed order cross section below $p_\perp\approx30$ GeV.

\EPSFIGURE[p]{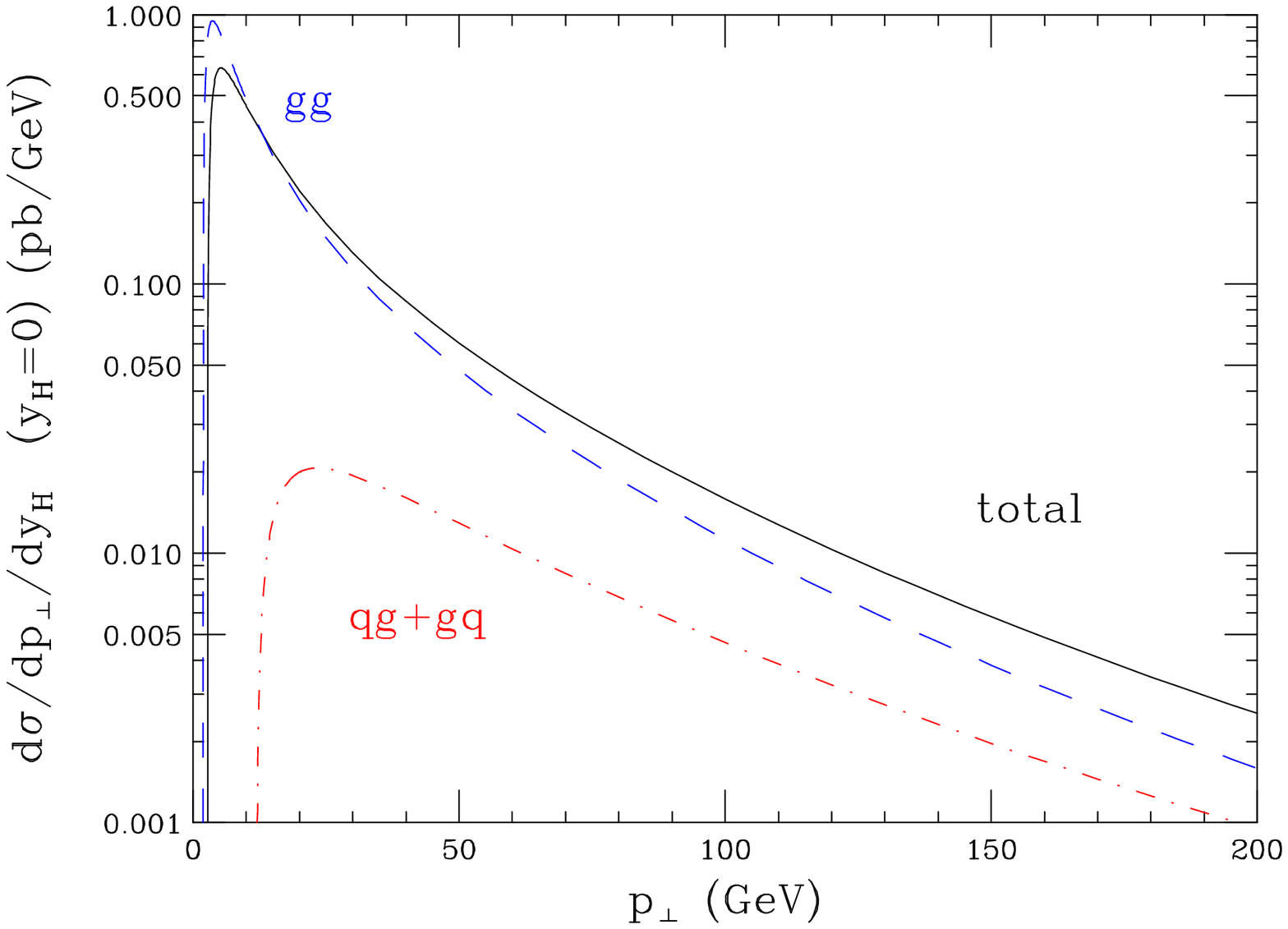, width=0.8\textwidth}
{The Higgs $p_\perp$ spectrum for $y_H=0$ at NLO in the
$m_t=\infty$ effective theory, separated into its
initial-state partonic subcomponents: gluon-gluon (dashes),
(anti)quark-gluon (dotdash), and total (solid).  The 
(anti)quark-(anti)quark contribution is very small and 
negative 
over the most of the range of this plot.
\label{fig:ptsubcomponents}}

\EPSFIGURE[p]{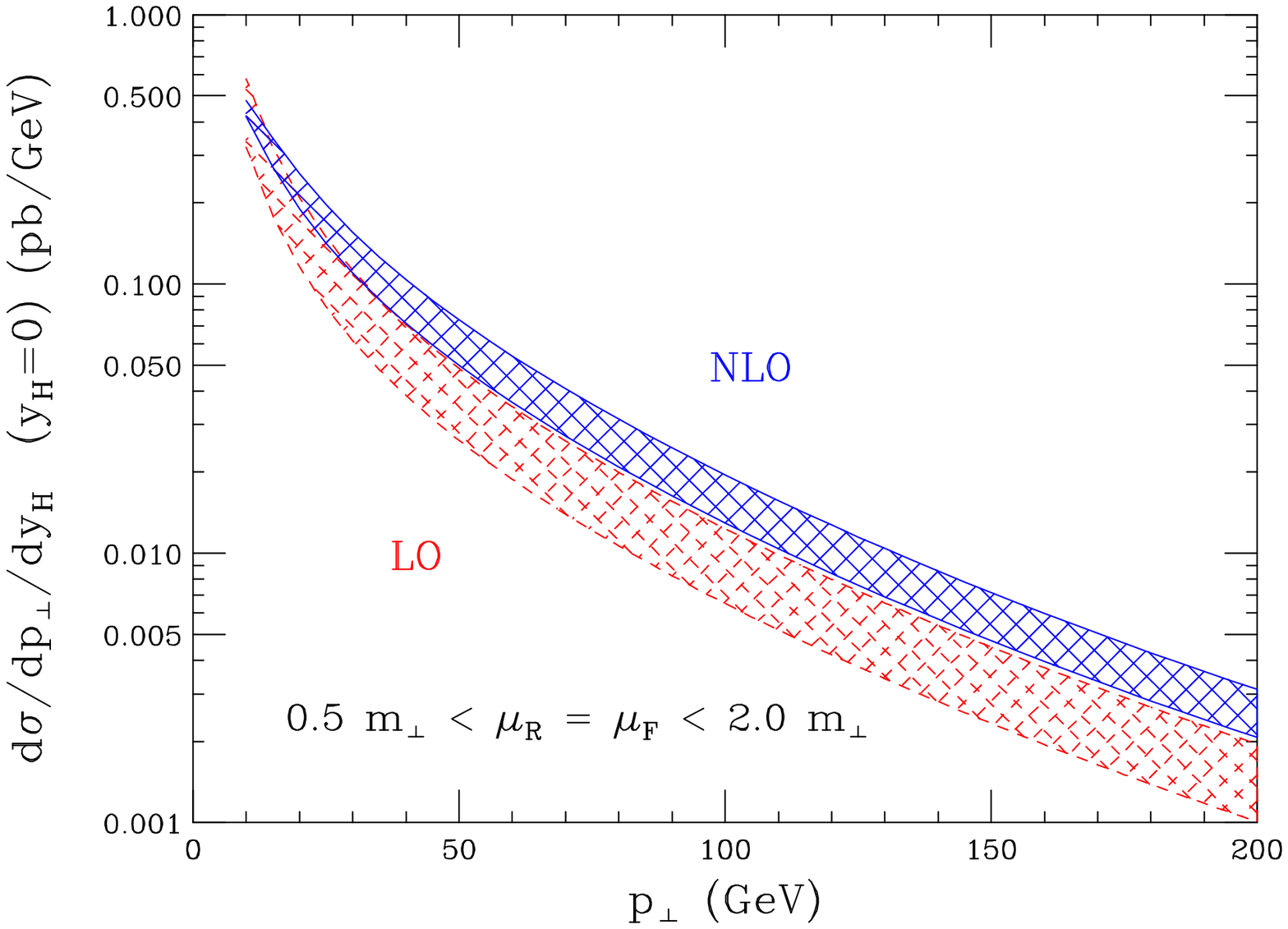, width=0.8\textwidth}
{The Higgs $p_\perp$ spectrum for $y_H=0$ at LO (dashes)
and NLO (solid), with the renormalization and factorization 
scales set equal $\mu_R=\mu_F$ and varied in the range 
$0.5m_\perp$ to $2m_\perp$.
\label{fig:ptbands}}

In Fig.~\ref{fig:ptScale} we plot the scale variation
of the differential cross section at the particular value of
transverse momentum $p_\perp=50$ GeV and rapidity $y_H=0$.
We plot both the diagonal variation $\mu_R=\mu_F=\chi m_\perp$
and the anti-diagonal variation $\mu_R=\chi m_\perp,\ \mu_F=m_\perp/\chi$,
with the cross section normalized to the value at $\mu_R=\mu_F=m_\perp$, 
($\chi=1$), and $0.1<\chi<10$.  For both the diagonal and the
anti-diagonal variation, the NLO curves are much less sensitive to 
the scale choice than the LO curves.  For instance, for a typical
range $0.5<\chi=\mu_R/m_\perp=\mu_F/m_\perp<2$, the LO ratio
varies from 0.74 to 1.38, whereas the NLO
ratio only varies from 0.83 to 1.22.
In Fig.~\ref{fig:ptScale} we also see that
the scale variation is least when the renormalization and factorization
scales are varied together with $\mu_R=\mu_F$.

\EPSFIGURE[p]{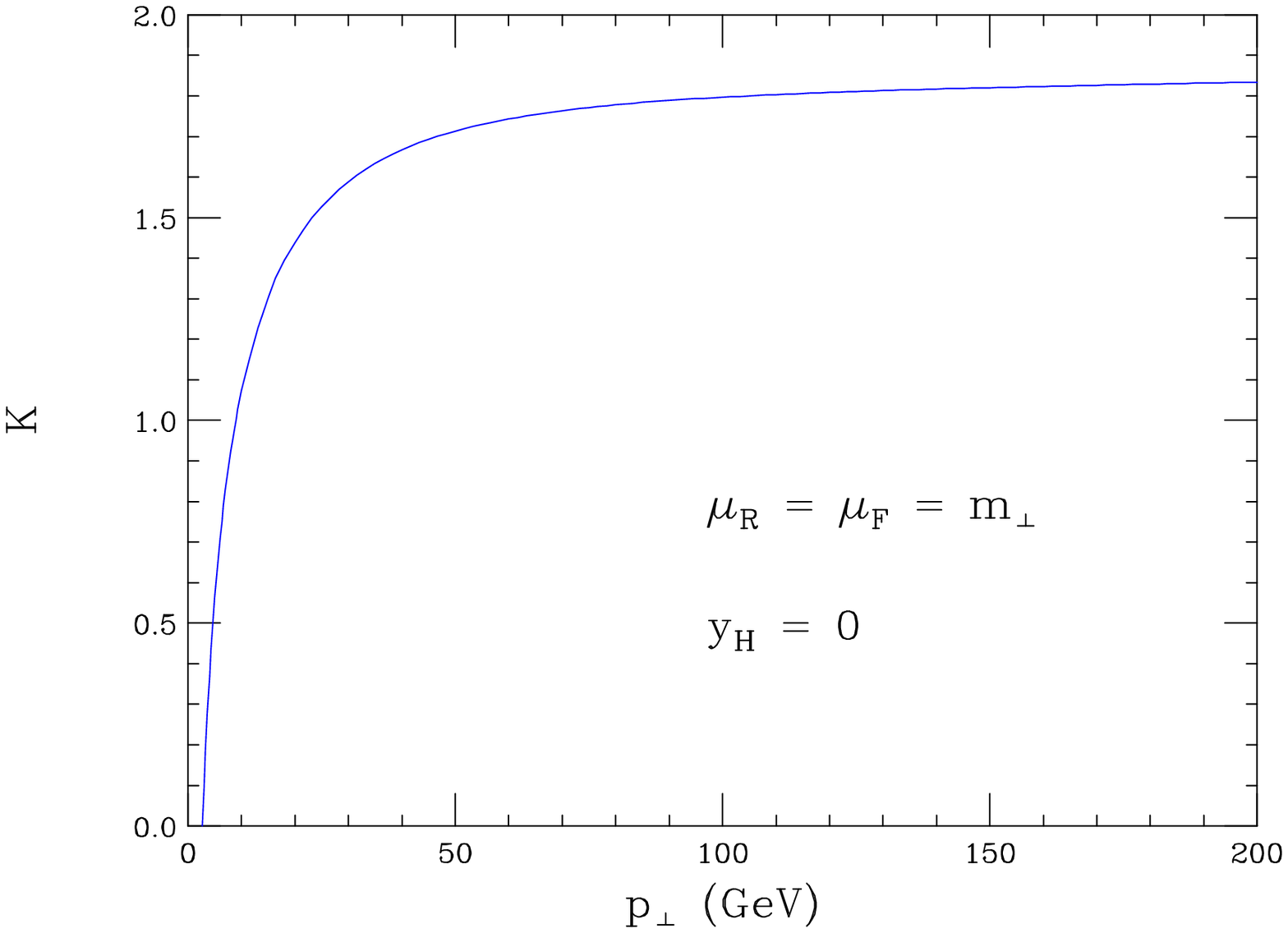, width=0.8\textwidth}
{The $K$-factor for the Higgs $p_\perp$ spectrum at
$y_H=0$.  The scales are set to $\mu_R=\mu_F=m_\perp$.
\label{fig:ptKfactor}}

\EPSFIGURE[p]{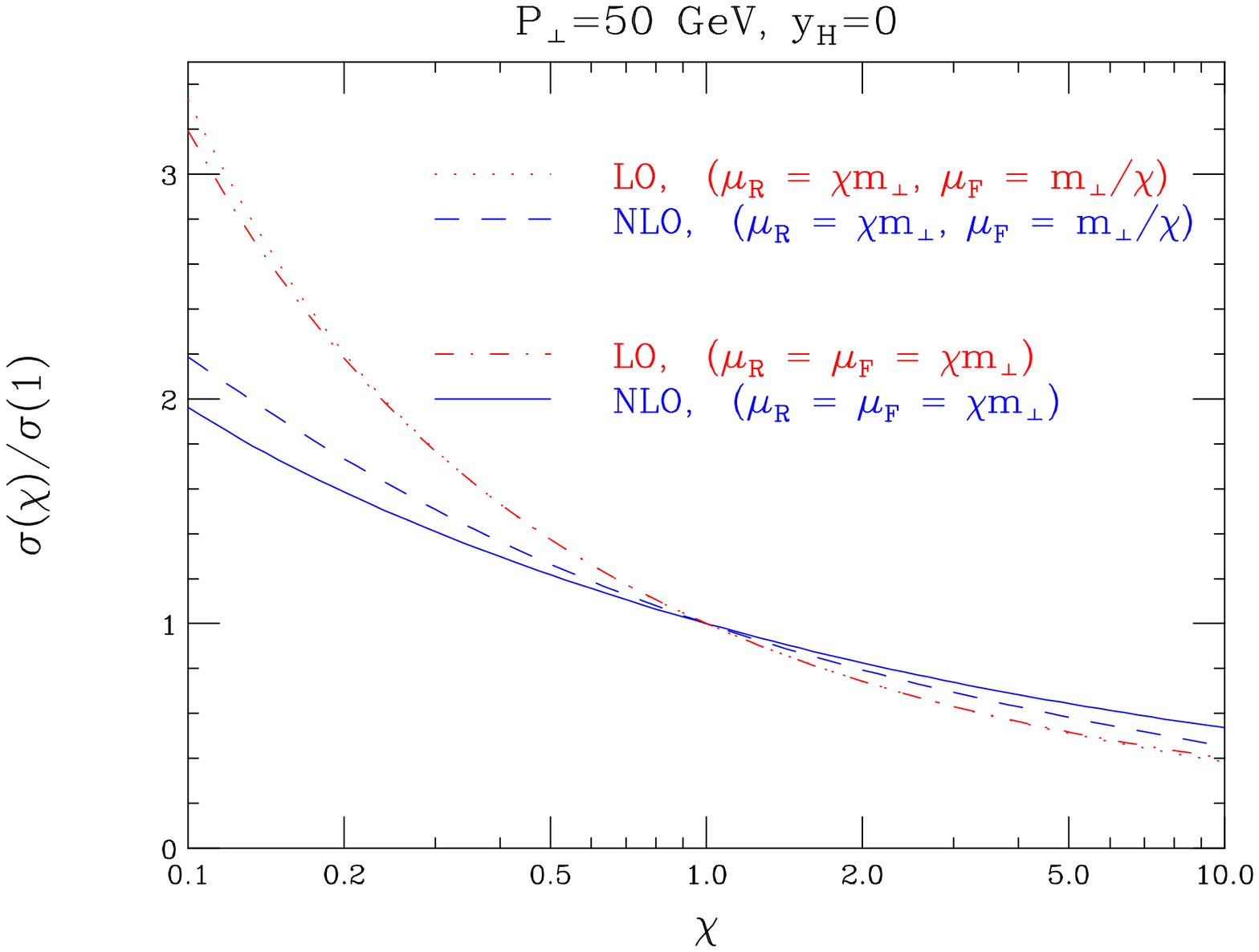, width=0.8\textwidth}
{The Higgs cross section for $y_H=0$ and $p_\perp=50$ GeV
as a function of varying renormalization $\mu_R$ and factorization
$\mu_F$ scales, normalized to the cross section at $\mu_R=\mu_F=m_\perp$.
The scales are varied together
at LO (dotdash) and NLO (solid)
and also inversely at LO (dots) and NLO (dashes).
\label{fig:ptScale}}

In addition to the $p_\perp$ spectrum at fixed $y_H$, 
we can treat the differential 
cross section as a rapidity distribution for fixed $p_\perp$.
This is plotted in Fig.~\ref{fig:rapidity} for $p_\perp=50$ and 100
GeV, at both LO and NLO.  Again we see the steep fall off for large
rapidities, due to the restriction of the available phase space,
and we see the increase of the NLO cross section over LO.
In Fig.~\ref{fig:YKfactor} we plot the $K$-factor for $p_\perp=50$ as
a function of the Higgs rapidity $y_H$.  It is very flat
as a function of $y_H$, except at large rapidities, where
it drops slightly, from 1.7 to about 1.6 at $|y_H|=4$.

As we have seen, the NLO corrections give a $K$-factor of about 
1.6 to 1.8 in the range of applicability of this fixed order
calculation, and this $K$ factor is fairly independent of the
exact values of $p_\perp$ and $y_H$.  In addition, the standard
variation of scales between $0.5m_\perp<\mu_R=\mu_F<2m_\perp$ suggests
an uncertainty in this NLO calculation on the order of about 20 percent.
One may also wonder what the uncertainty on this cross section is
due to the parton density functions.  We have calculated the
NLO $p_\perp$ and $y_H$ distributions using the MRST99 set 1 
distribution~\cite{MRST99} and found that the difference is less
than 4 percent over the full range of 30 GeV$<p_\perp<150$ GeV and
$|y_H|<2.5$.  If we restrict to $y_H=0$, the deviation is never
more than 2 percent.  This is in contradiction to the results of
Ref.~\cite{RSvN}, where they found a much larger variation, albeit
using the older PDFs of MRST98~\cite{MRST98}, CTEQ4~\cite{CTEQ4},
and GRV98~\cite{GRV98}.  Presumably, the more recent global analyses
produced less variability in their standard PDFs.
We expect that the differences resulting from the use of
the more recent CTEQ6M~\cite{CTEQ6M} and MRST2001~\cite{MRST2001}
will also be small, since neither update has a large change
in the region of $x$ and $Q^2$ gluon distributions
relevant to our calculation.  However, this does not necessarily 
mean that the uncertainty due to the PDFs is so small.  
The recent analysis on PDF uncertainties in~\cite{CTEQ6M} suggests
that the expected error on the gluon luminosity function in the
range of relevant $\hat s$ at the LHC is on the order of 5 percent.

\EPSFIGURE[p]{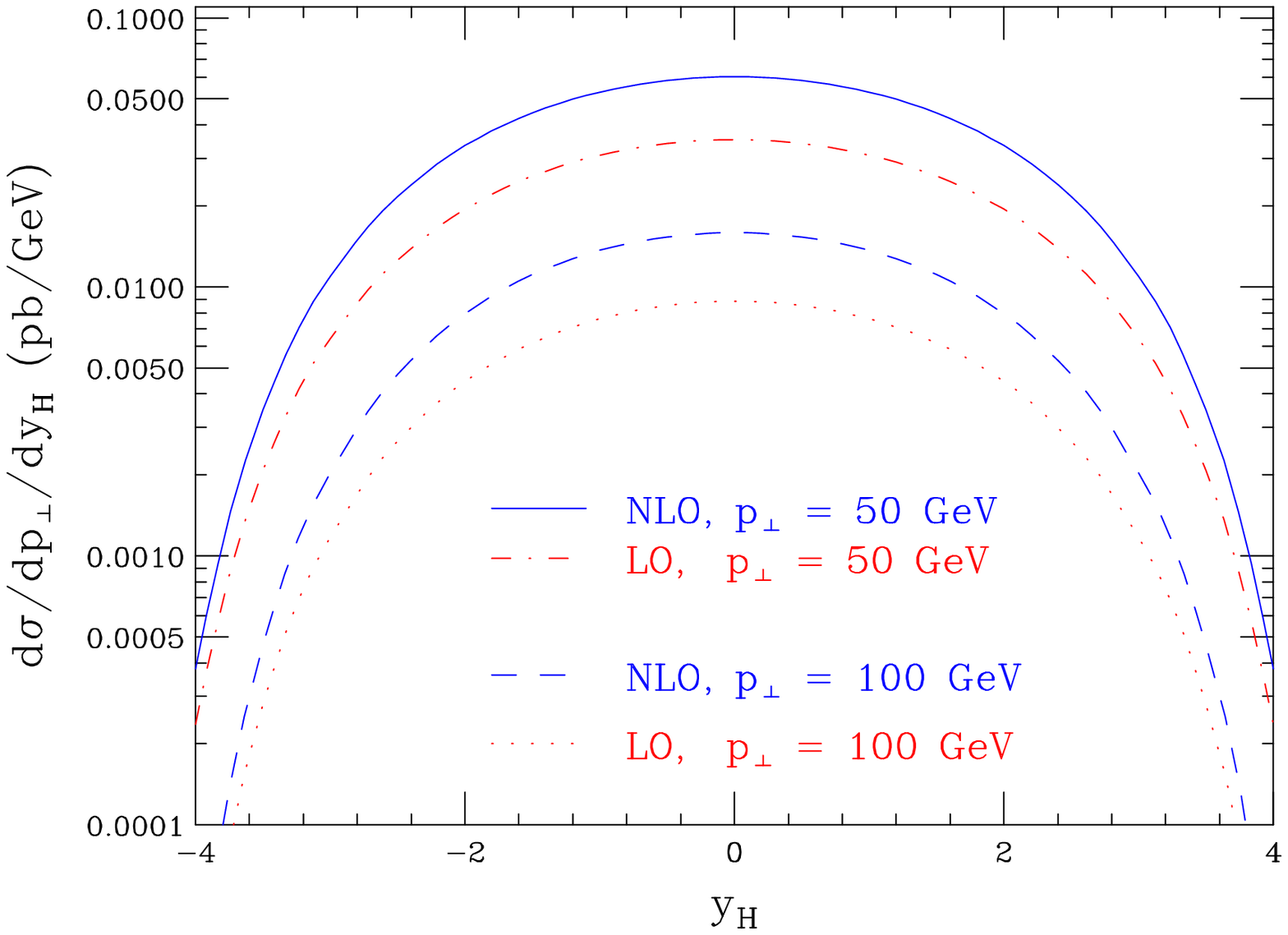, width=0.8\textwidth}
{The rapidity dependence of the Higgs cross section for two
different transverse momenta. The curves are
for $p_\perp=50$ GeV at NLO (solid) and LO (dotdash) and
for $p_\perp=100$ GeV at NLO (dashes) and LO (dots).
\label{fig:rapidity}}

\EPSFIGURE[p]{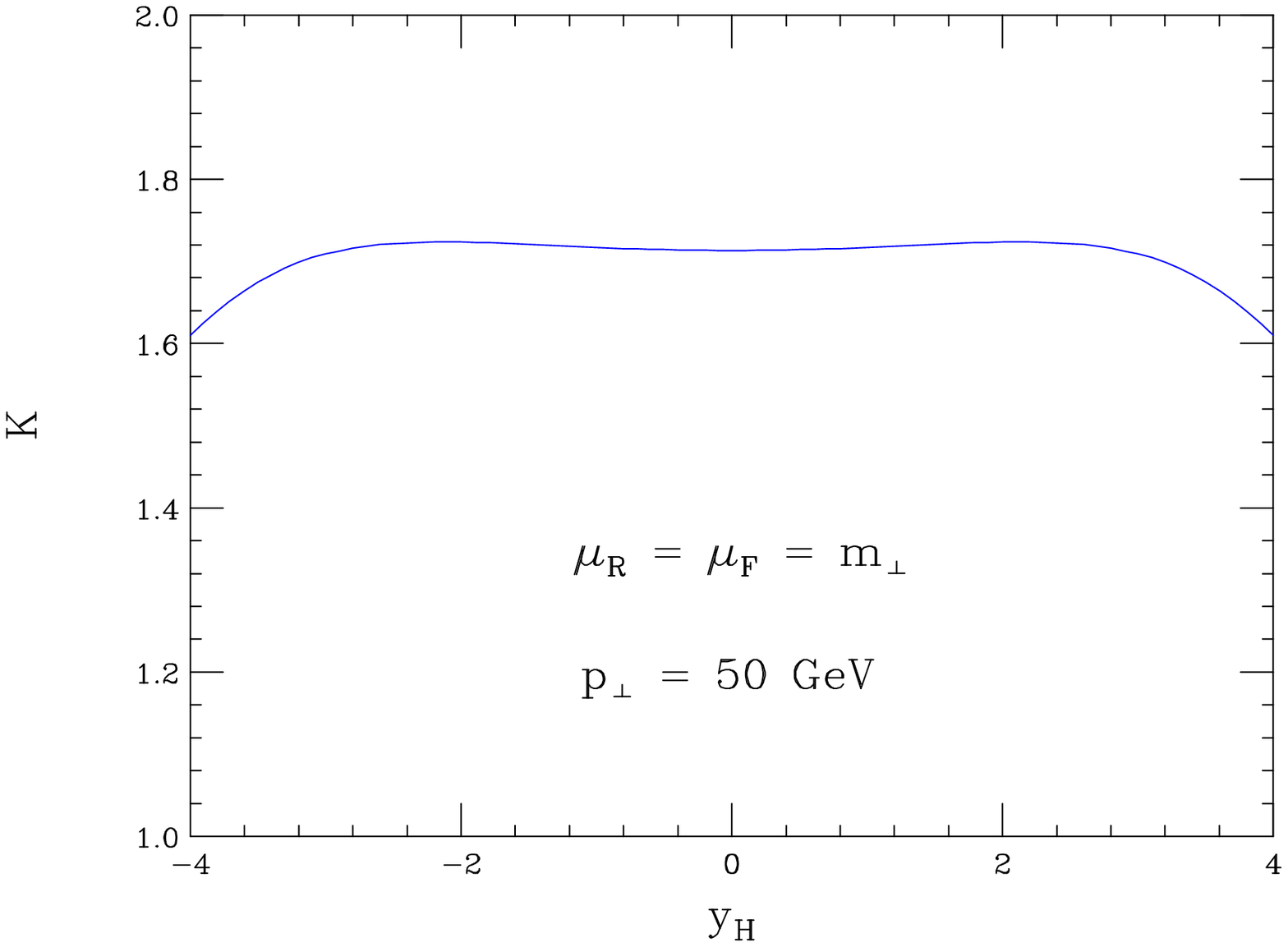, width=0.8\textwidth}
{The $K$-factor as a function of rapidity for
$p_\perp=50$ GeV.  The scales are set to $\mu_R=\mu_F=m_\perp$.
\label{fig:YKfactor}}


\section{Small-$p_\perp$ limit}
\label{sec:smallpt}

At small $p_\perp\ll m_H$, the $p_\perp$ spectrum becomes
unstable at any fixed order, due to large logarithms.  
At LO it diverges to positive infinity as $p_\perp\rightarrow0$, 
as seen in Fig.~\ref{fig:locomparison}, while at NLO it diverges to
negative infinity, as seen in Fig.~\ref{fig:ptY}.
These logarithmic corrections can be resummed, however, resulting
in a well-behaved physical $p_\perp$ spectrum, even for $p_\perp\ll m_H$.
In this section we compare our NLO results with what is expected
from the resummed cross section.

Following Collins, Soper, and Sterman \cite{CSS}, we can write the resummed
Higgs $p_\perp$ spectrum as an integral over an impact parameter:
\begin{equation}
	{d\sigma\over dp_{\perp}^{2}dy_{H}}(\mbox{resummed})
	\ =\  {m_H^2\,\sigma_0\bigl(\alpha_s(m_H)\bigr)\over2s}\int_0^\infty b\, db\, J_0(bp_\perp)\,W(b)
\ ,
\label{eq:resumm}
\end{equation}
with
\begin{eqnarray}
W(b)& =& \Bigl(C_{gi}\bigl(\alpha_s(b_0/b)\bigr)\circ f_i\Bigr)(\bar x_a^0;b_0/b)\,
\Bigl(C_{gj}\bigl(\alpha_s(b_0/b)\bigr)\circ f_j\Bigr)(\bar x_b^0;b_0/b)\nonumber\\
&&\times
\exp\left\{-\int_{b_0^2/b^2}^{m_H^2}
{dq^2\over q^2}\left[A\bigl(\alpha_s(q)\bigr)\ln{m_H^2\over q^2}+
B\bigl(\alpha_s(q)\bigr)\right]\right\}\ ,\label{eq:sudakov}
\end{eqnarray}
where $i$ and $j$ are implicitly summed over all massless partons
($g,q_f,\bar q_f$), and $b_0=2e^{-\gamma_E}$,
with Euler's constant $\gamma_E$.  
In eq.~(\ref{eq:sudakov}) we have introduced the
symbol $\circ$ to denote convolution, defined by
\begin{equation}
	\Bigl(f\circ g\Bigr)(x)\ =\ 
	\int_{x}^{1}{dz\over z}f(z)g(x/z)\ .
	\label{eq:conv}
\end{equation}
In eq.~(\ref{eq:sudakov}) the convolutions are evaluated at
\begin{eqnarray}
	{\bar x}_{a}^{0} & = & {m_{H}e^{y_{H}}\over\sqrt{s}}
	\nonumber  \\
	{\bar x}_{b}^{0} & = & {m_{H}e^{-y_{H}}\over\sqrt{s}}\ ,
	\label{eq:xbars}
\end{eqnarray}
and the factorization scale in the PDFs is set to $b_0/b$.
The parameters $A(\alpha_s)$, $B(\alpha_s)$, and the convolution
functions $C_{ij}(\alpha_s,z)$ can be expanded as a power series
in $\alpha_s$:
\begin{eqnarray}
A(\alpha_s)&=&\sum_{n=1}^\infty\left({\alpha_s\over2\pi}\right)^n\,A^{(n)}\ ,
\nonumber\\
B(\alpha_s)&=&\sum_{n=1}^\infty\left({\alpha_s\over2\pi}\right)^n\,B^{(n)}\ ,
\nonumber\\
C_{ij}(\alpha_s,z)&=&\delta_{ij}\delta(1-z)+\sum_{n=1}^\infty\left({\alpha_s\over2\pi}\right)^n\,C_{ij}^{(n)}(z)\ .
\end{eqnarray}
The known coefficients and functions in the power series (relevant to
Higgs production) are~\cite{Catani:vd,Hresum,dFG}
\begin{eqnarray}
A^{(1)}&=&2N_c\ ,\nonumber\\
A^{(2)}&=&2N_c\left[\left({67\over18}N_c-{5\over9}n_f\right)-{\pi^2\over6}N_c
\right]\ ,\nonumber\\
B^{(1)}&=&-2\beta_0\ ,\nonumber\\
B^{(2)}&=& -2\delta P^{(2)}_{gg}+\beta_0\left(\Delta
+{4\pi^2\over3}N_c\right)\ ,\label{eq:params}\\
C^{(1)}_{gg}(z)& =& \left({\Delta\over2}+{\pi^2\over2}N_c\right)\delta(1-z)\ ,\nonumber\\
C^{(1)}_{gq}(z)&=&C^\epsilon_{gq}(z)\ =\ C_Fz\ ,\nonumber
\end{eqnarray}
where
\begin{equation}
\delta P^{(2)}_{gg}\ =\ N_c^2\left({8\over3}+3\zeta(3)\right)-{1\over2}n_fC_F
-{2\over3}n_fN_c
\end{equation}
is the coefficient of the $\delta(1-z)$ term in $P^{(2)}_{gg}(z)$, the
NLO splitting function, and $\zeta(n)$ is the Riemann zeta function with
$\zeta(3)=1.202057\dots$.  The comparable coefficients in Drell-Yan
production have been given in Ref.~\cite{davies}.
All of the coefficients and functions in
eq.~(\ref{eq:params}), except $B^{(2)}$ and $C^{(1)}_{gg}(z)$, 
are universal, in the sense that they only depend on the color charges
of the initial-state partons involved in the scattering ($gg$ for
the present process), and not on
the final-state that is produced.  On the other hand, the expressions for
$C^{(1)}_{gg}(z)$ and $B^{(2)}$ given above are specific to Higgs production.
Of these coefficients, $B^{(2)}$ was obtained most recently in Ref.~\cite{dFG},
using the universality of the real emission contributions at small
$p_\perp$, combined with knowledge of the virtual correction 
amplitudes in the soft and collinear limits.   The universality structure
of all of these coefficients was revealed in more detail in 
Ref.~\cite{CdFG2}, where it was shown that the general resummation 
formula, eq.~(\ref{eq:resumm}), can be rewritten in such a way to make all
of $A(\alpha_s)$, $B(\alpha_s)$, and $C_{ij}(\alpha_s,z)$ universal,
by extracting the ``non-universal parts'' into a single new 
process-dependent parameter $H(\alpha_s)$.

The resummed formula, eq.~(\ref{eq:resumm}), can be expanded as a
power series in $\alpha_s$ for $0<p_\perp\ll m_H$, as in 
Ref.~\cite{ak}.  At NLO this yields
\begin{equation}
	{d\sigma\over dp_{\perp}^{2}dy_{H}}\Biggr|_{p_{\perp}\ll m_H}
\ =\
	{\sigma_{0}\over s}{m_{H}^{2}\over p_{\perp}^{2}}
\left[
\sum_{m=1}^2\sum_{n=0}^{2m-1}\left(
	{\alpha_{s}\over2\pi}\right)^{m}
	\,_{m}C_{n}\,\left(\ln{m_{H}^{2}\over 
	 p_{\perp}^{2}}\right)^{n}
+{\cal O}(\alpha_s^3)\right]
\ ,\label{eq:smallpt}
\end{equation}
where the coefficients, $_mC_n$, depend on the coefficients and functions
given in eq.~(\ref{eq:params}).  The formulae for the $_mC_n$ are given
explicitly in appendix~\ref{app:smallpt} in eq.~(\ref{eq:mCn}).
In particular the coefficient $_2C_0$ depends on $B^{(2)}$.
We have checked analytically that the small-$p_\perp$ limit of our 
calculation agrees exactly with eq.~(\ref{eq:smallpt}).  Thus,
we have verified the formula for $B^{(2)}$ for $gg\rightarrow H+X$
given in eq.~(\ref{eq:params}).  Some of the details involved in taking
the small-$p_\perp$ limit of our cross section are given in 
appendix~\ref{app:smallpt}.

In Fig.~\ref{fig:smallpt} we plot a comparison of our calculation at
both LO and NLO versus the corresponding small-$p_\perp$ limit formulae,
eq.~(\ref{eq:smallpt}).
As can be seen from the figure, the small-$p_\perp$ limit works very well
for $p_\perp\lesssim10$ GeV, and it is indistinguishable 
on the plot from the exact fixed order calculation for $p_\perp\lesssim5$ GeV.

In Ref.~\cite{RSvN} the soft-plus-virtual (S+V) approximation, which
retains all of the singular terms of the cross section as $z_i\rightarrow1$,
was considered as an estimate of the full NLO cross section.  
This approximation had previously been used to estimate the
NNLO corrections to the total Higgs production cross section~\cite{CdFG}.
In \cite{RSvN} it was
seen that this approximation is reasonable for larger $p_\perp$,
where the phase space for non-soft emissions is constricted.
In this vein we may consider our ``singular'' contributions 
from eqs.~(\ref{eq:ggsing}), (\ref{eq:gqsing}), (\ref{eq:qiaising}),
and (\ref{eq:qiajsing}) to be
a combined S+V and small-$p_\perp$ approximation, as it contains all
of the dominant soft-plus-virtual terms, and it contains all of the
leading contributions at small $p_\perp$ except the $_2C_0$ terms.
In addition, we note that it also contains the contribution of all 
collinear emissions, including those not singular as $z_i\rightarrow1$.
Because it contains the terms which dominate at large $p_\perp$, as
well as those that dominate at small $p_\perp$, one may hope that this
approximation will be good in the intermediate range of $p_\perp$ as
well.

In analogy to Ref.~\cite{RSvN} we define the ratio
\begin{equation}
R^\mathrm{sing}\ =\ {\left(d\sigma/dp_\perp^2/dy_H\right)
(\mbox{``singular'' only})\over
\left(d\sigma/dp_\perp^2/dy_H\right)(\mbox{exact})}\ .
\end{equation}
We plot the ratio $R^\mathrm{sing}$ in Fig.~\ref{fig:Rsing}
for two choices of the Higgs boson mass.  We also plot 
the 
similarly-defined ratio $R^\mathrm{S+V}$, where we have calculated the cross
section with soft-plus-virtual terms only, by keeping only those
terms in eqs.~(\ref{eq:ggsing}), (\ref{eq:gqsing}), (\ref{eq:qiaising}),
and (\ref{eq:qiajsing}) that are multiplied by $\delta(Q^2)$ or
a $+$function singularity, and setting $Q^2=0$ in all factors that 
are multiplied by a $+$function.  Note that since we write the $+$functions
in terms of $z_{i}$ rather than $Q^2$, our definition differs
from \cite{RSvN} by terms that are 
nonsingular as $z_i\rightarrow1$.  From the 
figure, we see that the pure S+V approximation does an excellent job at
$p_\perp=200$ GeV for $m_H=120$ GeV, 
overestimating the exact NLO cross section
by 2.5\%,  while the ``singular'' approximation also does well,
underestimating by 6.7\%.  The surprisingly good agreement of the
S+V approximation
is somewhat fortuitous;  keeping some non-leading
dependence on $z_i$ typically changes the cross section by about 
$5-10$\% at $p_\perp=200$ GeV.  
At medium to lower $p_\perp$, the ``singular''
approximation does better than S+V.  For instance, at $p_\perp=100$ GeV
the ``singular'' approximation underestimates by 4.7\%, while the
S+V approximation overestimates by 11.6\%.  This suggests an 
explanation for the observation in Ref.~\cite{RSvN} that the S+V 
approximation does worse for larger $m_H$.  Although increasing $m_H$
does restrict the phase space for real emission, it also
makes the logarithm $\log{m_H/p_\perp}$ larger for a given $p_\perp$.  
Thus, the range in $p_\perp$ for which
the small-$p_\perp$ logarithms dominate increases with $m_H$.  
To illustrate this effect, we have also plotted the same ratios for 
$m_H=200$ GeV.  For this larger Higgs mass, the ``singular''
approximation improves, whereas the pure S+V approximation gets worse.

\EPSFIGURE[p]{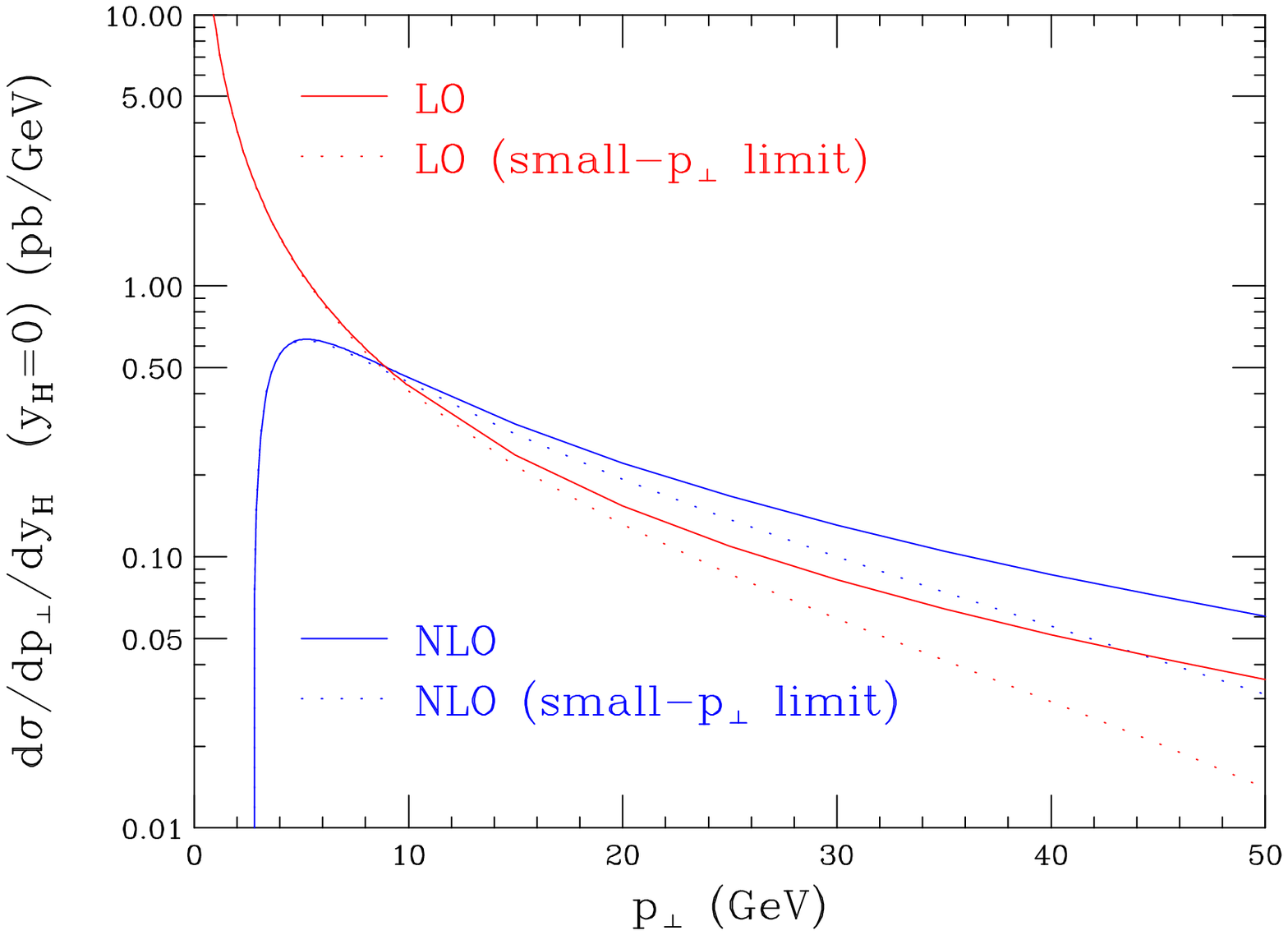, width=0.8\textwidth}
{The Higgs $p_\perp$ spectrum (solid curves) 
compared to the small-$p_\perp$
limit formulae Eq.~(\ref{eq:smallpt}) (dotted curves) at both LO and NLO.
All curves are at calculated at 
$y_H=0$ in the
$m_t=\infty$ effective theory.
\label{fig:smallpt}}

\EPSFIGURE[p]{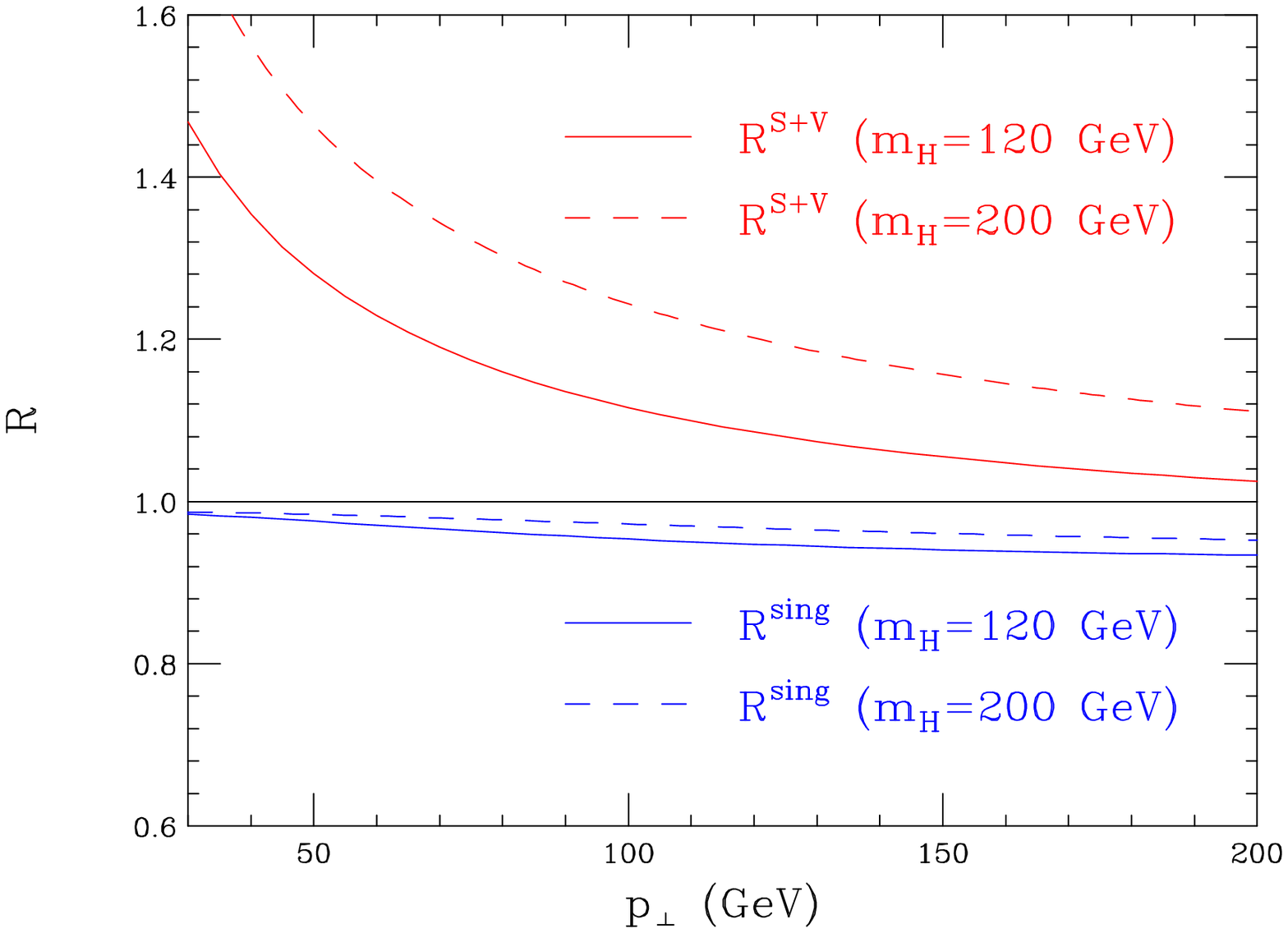, width=0.8\textwidth}
{The ratio of the NLO $p_\perp$ spectra, calculated with
either ``singular'' terms only or soft+virtual terms only, 
divided by the full NLO cross section.  The curves are for
$m_H=120$ GeV (solid) and $m_H=200$ GeV (dashes).
All cross sections are for $y_H=0$ in the
$m_t=\infty$ effective theory.
\label{fig:Rsing}}


\section{Conclusions}
\label{sec:conclusions}

In this paper we have presented a NLO calculation of the Higgs
boson transverse momenta and rapidity distributions, for nonzero 
Higgs $p_\perp$.  We have included all of the analytic formulae,
so that it is possible to directly investigate them in
various kinematic limits.
In particular, we have checked the small-$p_\perp$ limit, and
directly verified the formulae predicted from the resummation of
large logarithms of $\ln{m_H/p_\perp}$, including the $B^{(2)}$
coefficient.  We have also isolated the most important 
pieces of the cross section into the ``singular''
contributions, which include the soft+virtual contributions, the 
remaining collinear contributions, and most of the small-$p_\perp$
contributions.  This may be useful as a starting
point for a combined soft+virtual and small-$p_\perp$ resummation.

The numerical size of the NLO corrections are large, with a $K$-factor
rising slowly from 1.6 to 1.8 as the $p_\perp$ increases from 30 GeV
to 200 GeV.  The scale dependence is reduced at NLO, compared to 
LO, with a variation on the order of 20\% for 
$0.5m_\perp<\mu_R=\mu_F<2m_\perp$.  We expect that the theoretical
uncertainty from uncalculated higher orders (as well as due to the
$m_t\rightarrow\infty$ limit) is larger than that due to
uncertainties in the PDFs, although a detailed error analysis on
the effect of these uncertainties on this particular process
(as outlined in~\cite{CTEQ6M}) would be necessary to make this
definitive.

\vskip .2 cm

\section*{Acknowledgments}
 
We wish to thank Russel Kauffman, Tim Tait, Wu-Ki Tung,
and C.-P. Yuan for useful discussions.  C.G. would like to 
acknowledge the physics department at Saint John's
University (MN) for their hospitality during much of this project.
This work was supported by the US National
Science Foundation under grant PHY-0070443. 


\appendix

\section{Nonsingular Real Contributions}
\label{app:nsrc}

Here we give the ``nonsingular'' contributions coming from the real $H+2$ parton
cross section.  These terms are finite as $\epsilon\rightarrow0$ and as $Q^2\rightarrow0$, 
so that they can be written in 4 dimensions.  We have also removed 
any naive singularities in $p_\perp^2$, before the phase space integration of eq.~(\ref{eq:master}).  Of course, this separation into ``nonsingular'' and ``singular'' terms is not unambiguous, but the sum is definitely unique.  We find the separation useful, since the formulae for the ``nonsingular'' terms are very long, whereas the dominant contribution to the cross section comes from the ``singular'' terms given in the main text.  

\subsection{$gg\rightarrow H+X$ terms}

We give the result as a sum over terms coming from specific color-ordered 
helicity amplitudes (after subtracting from each the ``singular'' terms that contribute to 
$G^{({\rm 2R,s})}_{ij}$), plus
a correction term arising from the ${\cal O}(\epsilon)$ correction to the matrix element
times a collinear singularity, which is necessary to obtain the correct CDR result.  
That is, for the gluon-gluon ``nonsingular'' contribution we write
\begin{eqnarray}
G^{({\rm 2R, ns})}_{gg}&=& {1\over\hat s^2p_\perp^2Q^2}\Biggl\{N_c^2\Bigl(A_0+A_{(1234)}+A_{(2341)}+A_{(3412)}\nonumber\\
&&\qquad\qquad\qquad+A_{(4123)}+A_{(1324)}+A_{(2413)}+A_{(3241)}+A_\epsilon\Bigr)\label{eq:ggterms}\\
&&\qquad\qquad+n_fC_F\Bigl(B_{1(+-)}+B_{1(++)}
\Bigr)
+n_fN_c\Bigl(B_{2(+-)}+B_{2(++)}
\Bigr)\Biggr\}\nonumber\ .
\end{eqnarray}
For definiteness, $A_0$ comes from $Hgggg$ squared amplitudes with helicities $(++++)$, $(----)$, or $(++--)$, where all helicities are taken to be for out-going momenta.  The $A_{(ijkl)}$ come from the $Hgggg$ color-ordered amplitude $m(i-,j+,k+,l+)$ squared, as well as all other squared amplitudes that are identical by color-reversal, helicity-reversal, or exchange of the final-state gluons.  For example, $A_{(3412)}$ comes from $m(3-,4+,1+,2+)$ plus all identical color-ordered amplitudes, where 1 and 2 refer to the incoming gluons.  (We use the cyclical symmetry of the color-ordered amplitudes to always put the negative helicity gluon in the first position.)
The terms, $B_{i(+-)}$ and $B_{i(++)}$, come from $Hggq\bar q$ squared amplitudes with unlike-helicity gluons and like-helicity gluons, respectively.  
Finally, $A_\epsilon$ comes from 
the ${\cal O}(\epsilon)$ corrections to the squared $Hgggg$ matrix element 
that are not proportional to $\delta(Q^2)$ and cannot be identified as
a contribution to the $C^\epsilon_{ij}$, and therefore
have not been included
in $G^{({\rm 2s})}_{gg}$, eq.~(\ref{eq:ggsing}).  
The $Hggq\bar q$ squared matrix element also has
${\cal O}(\epsilon)$ corrections, but they have all been included in $G^{({\rm 2s})}_{ij}$, eq.~(\ref{eq:ggsing}).

For later use in these formulae, we define
\begin{eqnarray}
L_{1i}&=& \ln{m_{H}^{2}\over {\hat s}z^{ 2}_{i}}\nonumber\\
L_{2i}&=& \ln{m_{H}^{2}{\hat s}\over(A-{\hat s}z_{i})^{2}}\\
L_{3}& =& \ln{A+B\over A-B}\ ,\nonumber
\label{eq:logs}
\end{eqnarray}
where
\begin{eqnarray}
	A&=& {\hat s}+m_{H}^{2}-Q^{2}\ ,\nonumber\\
	\label{eq:vhat}
	B&=& \left(A^{2}-4m_{H}^{2}{\hat s}\right)^{1/2}\ .
\end{eqnarray}
We obtain from the $Hgggg$ squared matrix elements:
\begin{eqnarray}
	A_0
	&=&\left[ ({\hat t}/z_{a})^{4}+({\hat u}/z_{b})^{4}\right]{p_\perp^2Q^2\over
	 Q_{\perp}^{4}}\left(
	5-7{Q^{2}\over Q_{\perp}^{2}}+{20\over3}{Q^{4}\over Q_{\perp}^{4}}
	\right)\nonumber\\
&&+\hat s^2Q^2p_\perp^2
	\left({17\over3}+4\ln{p_{\perp}^{2}/Q_{\perp}^{2}}\right)\,,\\
	A_{(1234)} & = & -{1\over2}\left[\left(({\hat s}
	p_{\perp}^{2}/{\hat t})^{4}+(m_{H}^{2}Q^{2}/{\hat t})^{4}\right)
	L_{1a}+	{\hat u}^{4}L_{1b}\right]\nonumber\\
	&  & 
	+{{\hat s}m_{H}^{4}Q^{2}{\hat u}^{3}\over2
	{\hat t}({\hat u}-m_{H}^{2})({\hat t}-m_{H}^{2})}
	\left(L_{1a}+L_{1b}\right)\nonumber\\
	&  & +{{\hat s}p_{\perp}^{2}Q^{2}{\hat u}^{3}\over2
	A({\hat t}-m_{H}^{2})}(L_{2b}-L_{1b})
	 +{{\hat s}p_{\perp}^{2}Q^{2}
	 \left(m_{H}^{8}+({\hat u}-m_{H}^{2})^{4}\right)\over2
	A{\hat t}({\hat u}-m_{H}^{2})}
	(L_{2a}-L_{1a})\nonumber\\
	 &  & +{\hat s}p_{\perp}^{2}m_{H}^{2}Q^{2}\left[
	 {{\hat u}^{4}\over2B^{2}{\hat t}^{2}}+
	 {{\hat u}^{2}\over2B^{2}}\right]
	 +\left({\hat s}p_{\perp}^{2}m_{H}^{2}Q^{2}\right)^{2}
	 \left[{-6\over B^{4}}-{4\over{\hat t}^{4}}+{8\over
	 B^{2}{\hat t}^{2}}\right]\nonumber\\
	 &  & + L_{3} \left\{
	 {{\hat s}p_{\perp}^{2}{\hat u}^{3}({\hat u}+{\hat t})
	 \over B{\hat t}}+\left({\hat s}p_{\perp}^{2}m_{H}^{2}Q^{2}\right)^{2}
	 \left[{3A\over B^{5}}-{1\over AB^{3}}\right]\right.\\
	 &  &\qquad -{\hat s}p_{\perp}^{2}m_{H}^{2}Q^{2}\left[
	 {1\over B{\hat t}}\left({\hat t}^{2}+{\hat t}{\hat u}+4{\hat 
	 u}^{2}-2m_{H}^{2}Q^{2}\right)\right.\nonumber\\
	 & &\qquad\qquad\qquad\quad\left.+
	 {A\over2B^{3}}\left({\hat t}^{2}+3{\hat t}{\hat u}
	 +3{\hat u}^{2}-6m_{H}^{2}Q^{2}\right)\right.\nonumber\\
	 & &\qquad\qquad\qquad\quad\left.\left.+{1\over2AB}\left(
	 {\hat t}^{2}+{\hat t}{\hat u}+7{\hat u}^{2}
	 -2m_{H}^{2}Q^{2}\right)\right]\right\}\ ,\nonumber
	\label{eq:R2}\\
A_{(2341)}&=& A_{(1234)}\quad\mbox{with}\quad\Bigl\{(\hat t,a)\leftrightarrow(\hat u,b)\Bigr\}\ ,
\\
	A_{(3412)} & = & {{\hat s}p_{\perp}^{2}Q^{2}A^{3}\over2{\hat t}
	({\hat u}-m_{H}^{2})}(L_{2a}+L_{1b})
	+{{\hat s}p_{\perp}^{2}({\hat u}+{\hat t})\over16{\hat u}{\hat t}B}
	\left[A^{4}+6A^{2}B^{2}+B^{4}\right]L_{3}
	\nonumber\\
	&&+\Biggl\{
-{{\hat s}p_{\perp}^{2}\over2{\hat u}{\hat t}}\left(
	\left({\hat s}-Q^{2}\right)^{4}+m_{H}^{8}
+2Q^2A\left({\hat s}-Q^{2}\right)^{2}-2Q^2m_{H}^{6}\right)
\nonumber\\
	&&\qquad
	-{\left({\hat s}p_{\perp}^{2}m_{H}^{2}Q^{2}\right)^{2}
	\over{\hat u}^{4}}+{2{\hat s}p_{\perp}^{2}m_{H}^{2}Q^{2}(A^2-\hat sm_H^2)\over{\hat 
	u}^{2}}\Biggr\}L_{1b}\nonumber\\
	&&+{{\hat s}p_{\perp}^{2}(Q^{2}-{\hat u})^{3}\over8{\hat u}
	{\hat t}(Q^{2}-{\hat t})}
	\left((Q^{2}-{\hat t}){\hat 
	s}+Q^{2}{\hat u}\right)\left[{4\over3}+2{p_{\perp}^{2}\over Q_{\perp}^{2}}
	+4{p_{\perp}^{4}\over Q_{\perp}^{4}}-{44\over3}
	{p_{\perp}^{6}\over Q_{\perp}^{6}}\right]\nonumber\\
	&&+{{\hat s}p_{\perp}^{2}(Q^{2}-{\hat t})^{3}\over8{\hat u}
	{\hat t}(Q^{2}-{\hat u})}
	\left((Q^{2}-{\hat u}){\hat 
	s}+Q^{2}{\hat t}\right)\left[{4\over3}+2{p_{\perp}^{2}\over Q_{\perp}^{2}}
	+4{p_{\perp}^{4}\over Q_{\perp}^{4}}-{44\over3}
	{p_{\perp}^{6}\over Q_{\perp}^{6}}\right]\nonumber\\	
	&&+{{\hat s}p_{\perp}^{2}(Q^{2}-{\hat u})^{2}\over4{\hat u}
	{\hat t}(Q^{2}-{\hat t})}
	\Biggl[-3({\hat t}-m_{H}^{2})\left((Q^{2}-{\hat t}){\hat 
	s}+Q^{2}{\hat u}\right)\nonumber\\
	&&\qquad\qquad-Q^{2}\left(m_{H}^{2}({\hat t}-m_{H}^{2})
	+Q^{2}({\hat u}-m_{H}^{2})\right)\Biggr]\left[
	1+2{p_{\perp}^{2}\over Q_{\perp}^{2}}
	-6{p_{\perp}^{4}\over Q_{\perp}^{4}}\right]\nonumber\\
	&&+{{\hat s}p_{\perp}^{2}(Q^{2}-{\hat t})^{2}\over4{\hat u}
	{\hat t}(Q^{2}-{\hat u})}
	\Biggl[-3({\hat u}-m_{H}^{2})\left((Q^{2}-{\hat u}){\hat 
	s}+Q^{2}{\hat t}\right)\nonumber\\
	&&\qquad\qquad+3Q^{2}\left(m_{H}^{2}({\hat u}-m_{H}^{2})
	+Q^{2}({\hat t}-m_{H}^{2})\right)\nonumber\\
	&&\qquad\qquad+4{\hat u}{\hat s}^{2}\Biggr]\left[
	1+2{p_{\perp}^{2}\over Q_{\perp}^{2}}
	-6{p_{\perp}^{4}\over Q_{\perp}^{4}}\right]   \label{eq:R3}\\
	&&+{{\hat s}p_{\perp}^{2}(Q^{2}-{\hat u})\over2{\hat u}
	{\hat t}(Q^{2}-{\hat t})}
	\Biggl[3({\hat t}-m_{H}^{2})^{2}\left((Q^{2}-{\hat t}){\hat 
	s}+Q^{2}{\hat u}\right)\nonumber\\
	&&\qquad\qquad+3({\hat t}-m_{H}^{2})
	Q^{2}\left(m_{H}^{2}({\hat t}-m_{H}^{2})
	+Q^{2}({\hat u}-m_{H}^{2})\right)\nonumber\\
	&&\qquad\qquad
	+Q^{2}{\hat u}\left(m_{H}^{2}({\hat t}-Q^{2})+Q^{2}({\hat u}-m_{H}^{2})
	\right)\Biggr]\left[
	1-2{p_{\perp}^{2}\over Q_{\perp}^{2}}\right]\nonumber\\
	&&+{{\hat s}p_{\perp}^{2}(Q^{2}-{\hat t})\over2{\hat u}
	{\hat t}(Q^{2}-{\hat u})}
	\Biggl[3({\hat u}-m_{H}^{2})^{2}\left((Q^{2}-{\hat u}){\hat 
	s}+Q^{2}{\hat t}\right)	+8{\hat u}{\hat t}{\hat s}^{2}+2{\hat u}{\hat s}^{3}
	\nonumber\\
	&&\qquad\qquad
	-2Q^{2}{\hat u}({\hat u}-Q^{2})^{2}
	-3m_{H}^{2}Q^{2}({\hat t}-m_{H}^{2})^{2}
	-3Q^{2}(m_{H}^{2}-Q^{2}){\hat t}^{2}\nonumber\\
	&&\qquad\qquad
	-Q^{2}{\hat u}\left(4{\hat u}{\hat t}-{\hat u}m_{H}^{2}
	-Q^{2}{\hat t}+2{\hat t}^{2}-4m_{H}^{4}\right)\nonumber\\
	&&\qquad\qquad
	+3m_{H}^{2}Q^{4}({\hat t}-m_{H}^{2})
	+m_{H}^{2}Q^{2}{\hat u}({\hat t}-Q^{2})\Biggr]
	\left[
	1-2{p_{\perp}^{2}\over Q_{\perp}^{2}}\right]\nonumber\\
	&&-{4\left({\hat s}p_{\perp}^{2}m_{H}^{2}Q^{2}\right)^{2}
	\over{\hat u}^{4}}
	+{{\hat s}p_{\perp}^{2}m_{H}^{2}Q^{2}B^{2}
	\over2{\hat u}^{2}}
	+{{\hat s}^{2}p_{\perp}^{2}m_{H}^{4}
	\over6}\left({({\hat s}+Q^{2})\over{\hat u}{\hat t}}+
	{Q^{2}\over{\hat u}^{2}}+{Q^{2}\over{\hat t}^{2}}\right)\nonumber\\
	&&
	+{2{\hat s}^{2}p_{\perp}^{2}Q^{2}m_{H}^{4}
	\over{\hat u}^{2}}
	+{{\hat s}^{2}p_{\perp}^{2}m_{H}^{4}
	\over{\hat u}}
	-{{\hat s}^{2}p_{\perp}^{2}	\over12{\hat u}{\hat t}}
	\left(30m_{H}^{6}+54Q^{4}m_{H}^{2}+8Q^{6}\right)
	\nonumber\\
	&&
	+{{\hat s}p_{\perp}^{2}	\over12{\hat u}{\hat t}}
	\biggl[11{\hat s}^{4}+17m_{H}^{8}
	+Q^{2}(61{\hat u}^{2}{\hat t}+17{\hat u}^{3}+73{\hat u}{\hat t}^{2}
	+29{\hat t}^{3})\nonumber\\
	&&\qquad\quad
	+m_{H}^{2}(24{\hat u}^{2}{\hat t}+6{\hat u}^{3}+36{\hat u}{\hat t}^{2}
	+18{\hat t}^{3})
	+Q^{4}(-21{\hat u}^{2}-33{\hat t}^{2}-52{\hat u}{\hat t})\nonumber\\
	&&\qquad\quad
	+m_{H}^{2}Q^{2}(-73{\hat u}^{2}-109{\hat t}^{2}-170{\hat u}{\hat t})
	+m_{H}^{4}(-23{\hat u}^{2}-35{\hat t}^{2}-52{\hat u}{\hat t})
	\nonumber\\
	&&\qquad\quad+
	m_{H}^{4}Q^{2}(134{\hat t}+110{\hat u})+4Q^{8}+52m_{H}^{2}Q^{6}
	+20m_{H}^{4}Q^{4}-22m_{H}^{6}Q^{2}
	\biggr]\ ,\nonumber	\\
A_{(4123)}&=& A_{(3412)}\quad\mbox{with}\quad\Bigl\{(\hat t,a)\leftrightarrow(\hat u,b)\Bigr\}\ ,
\\
	A_{(1324)} & = & -{1\over2}\left[\left(({\hat s}
	p_{\perp}^{2}/{\hat t})^{4}+(m_{H}^{2}Q^{2}/{\hat t})^{4}\right)
	L_{1a}+	{\hat u}^{4}L_{1b}\right]
	+{{\hat s}^{2}p_{\perp}^{2}m_{H}^{4}Q^{2}\over{\hat t}^{2}}
	L_{1a}\nonumber\\
	&  & 
	+\left[{{\hat s}m_{H}^{4}Q^{2}{\hat u}^{3}\over2
	{\hat t}({\hat u}-m_{H}^{2})({\hat t}-m_{H}^{2})}
	+{{\hat s}p_{\perp}^{2}{\hat u}^{3}\over2{\hat t}}\right]
	\left(L_{1a}+L_{1b}\right)\nonumber\\
	&  & +{{\hat s}^{2}p_{\perp}^{2}(1-z_{b}){\hat u}^{3}\over2
	A({\hat t}-m_{H}^{2})}(L_{2b}-L_{1b})
	 +{{\hat s}^{2}p_{\perp}^{2}(1-z_{a})
	 \left(m_{H}^{8}+({\hat u}-m_{H}^{2})^{4}\right)\over2
	A{\hat t}({\hat u}-m_{H}^{2})}
	(L_{2a}-L_{1a})\nonumber\\
	 &  & +{{\hat s}^{2}p_{\perp}^{2}m_{H}^{4}Q^{2}\over AB}L_{3}
	 +{{\hat s}p_{\perp}^{2}m_{H}^{2}Q^{2}\over2{\hat t}^{4}}
	 \left[\left({\hat s}p_{\perp}^{2}\right)^{2}-6{\hat 
	 s}p_{\perp}^{2}m_{H}^{2}Q^{2}+m_{H}^{4}Q^{4}\right]\ ,
	\label{eq:R4}\\
A_{(2413)}&=& A_{(1324)}\quad\mbox{with}\quad\Bigl\{(\hat t,a)\leftrightarrow(\hat u,b)\Bigr\}\ ,
\\
	A_{(3241)} & = & 
	{{\hat s}^{2}p_{\perp}^{2}A^{3}(1-z_{a})\over2{\hat t}
	({\hat u}-m_{H}^{2})}(L_{2a}-L_{1a})
	\nonumber\\
	&&+\Biggl[
	-{\left({\hat s}p_{\perp}^{2}m_{H}^{2}Q^{2}\right)^{2}
	\over{\hat t}^{4}}
	+{{\hat s}p_{\perp}^{2}m_{H}^{4}Q^{4}\over{\hat u}{\hat 
	t}}-{{\hat s}p_{\perp}^{2}m_{H}^{2}Q^{2}A^{4}
	\over2{\hat u}{\hat t}({\hat u}-m_{H}^{2})({\hat t}-m_{H}^{2})}
	\nonumber\\
	&&\qquad	+{{\hat s}p_{\perp}^{2}Q^{2}m_{H}^{2}({\hat u}+{\hat 
	t})(2A^{2}-{\hat s}m_{H}^{2})\over{\hat u}{\hat t}^{2}} 
	\Biggr]L_{1a}\nonumber\\
	&&+{{\hat s}^{2}p_{\perp}^{2}Q^{2}(Q^{2}-{\hat u})^{2}
	\over 2{\hat u}{\hat t}(Q^{2}-{\hat t})^{2}}\left(
	-{\hat u}{\hat t}-(Q^{2}-{\hat t})^{2}\right)
	\left[-3+10{Q^{2}\over Q_{\perp}^{2}}-6{Q^{4}\over 
	Q_{\perp}^{4}}\right]\nonumber\\
	&&+{{\hat s}^{2}p_{\perp}^{2}Q^{2}(Q^{2}-{\hat u})
	\over {\hat u}{\hat t}(Q^{2}-{\hat t})^{2}}\left(
	{\hat u}{\hat t}(Q^{2}-{\hat u})-(Q^{2}-{\hat t})^{3}
	-m_{H}^{2}(Q^{2}-{\hat t})^{2}\right.\nonumber\\
	&&\qquad\qquad\left.
	-m_{H}^{2}(Q^{2}-{\hat t})
	(Q^{2}-{\hat u})\right)
	\left[-1+2{Q^{2}\over Q_{\perp}^{2}}\right]\nonumber\\
	&&+{\hat s}p_{\perp}^{2}m_{H}^{2}Q^{2}\left[
	{B^{2}\over2{\hat t}^{2}}-{2m_{H}^{2}Q^{2}\over{\hat t}^{2}}
	+{({\hat u}+{\hat t})^{2}\over2{\hat u}{\hat t}}\right]\nonumber\\
	&&-{4({\hat s}p_{\perp}^{2}m_{H}^{2}Q^{2})^{2}\over{\hat t}^{4}}
	+{{\hat s}^{2}p_{\perp}^{2}Q^{2}\over4{\hat u}{\hat t}}\left[
	({\hat t}+{\hat u})^{2}-({\hat t}+{\hat u})(6Q^{2}+4m_{H}^{2})
	+6Q^{4}+8m_{H}^{2}Q^{2}\right]\nonumber\\
	&&+{{\hat s}^{2}p_{\perp}^{2}Q^{2}m_{H}^{4}({\hat 
	t}+{\hat u})^{2}\over4{\hat u}^{2}{\hat t}^{2}}
	\ +\ \Bigl[(\hat t,a)\leftrightarrow(\hat u,b)\Bigr]
	\ ,\\
A_\epsilon&=& 4({\hat s}p_{\perp}^{2}m_{H}^{2}Q^{2})^{2}
	\left({1\over{\hat t}^{4}}+{1\over{\hat u}^{4}}\right)\ .
\end{eqnarray}

We obtain from the $Hggq\bar q$ squared matrix elements:
\begin{eqnarray}
	B_{1(+-)}
	& = & 
	{{\hat s}^{4}p_{\perp}^{2}z_{a}
	(1-z_{a})^{3}\over{\hat t}}
	+{{\hat s}^{4}p_{\perp}^{6}z^{3}_{a}(1-z_{a})\over{\hat 
	t}^{3}}+
	{4{\hat s}^4p_{\perp}^{4}z^{2}_a(1-z_a)^2\over 
        {\hat t}^2}
	\nonumber\\
	&&
	-{\hat s}^{2}p_{\perp}^{2}Q^{2}
	\left(1+\ln{p_{\perp}^{2}/Q_{\perp}^{2}}\right)
	\ +\ \Bigl[(\hat t,a)\leftrightarrow(\hat u,b)\Bigr]
	\ ,\label{eq:pm}\\
	B_{2(+-)}
	&=&{1\over3}\left({\hat t\over z_a}\right)^4
	{p_{\perp}^{2}\over Q_{\perp}^{2}}
	\left({p_{\perp}^{6}-Q^6-Q_\perp^6\over Q_{\perp}^{6}}
	\right)
	-{1\over3}{\hat s}^{2}p_{\perp}^{2}Q^{2}
	\ +\ \Bigl[(\hat t,a)\leftrightarrow(\hat u,b)\Bigr]
	\ ,\\
	B_{1(++)}
	& = & 
	{{\hat s}^{4}p_{\perp}^{2}z^{3}_{a}(1-z_{a})\over{\hat t}}
	+{{\hat s}^{4}p_{\perp}^{6}z_{a}(1-z_{a})^{3}\over{\hat 
	t}^{3}}+{4{\hat s}^4p_\perp^4z^{2}_{a}(1-z_a)^2\over {\hat t}^2}
	\nonumber\\
	&&-{{\hat s}^{2}p_{\perp}^{2}Q^{2}\over
	Q_{\perp}^{4}{\hat u}{\hat t}}
	\left(\left({\hat u}{\hat t}+p_{\perp}^{2}Q^{2}\right)^2
	+2\hat sp_{\perp}^{2}Q^{2}Q_\perp^2\right)
	\\
	&&+\,{{\hat s}^{2}p_{\perp}^{2}Q^{2}\over{\hat u}{\hat t}}
	\left({\hat s}^{2}+Q^{4}\right)\ln{Q_{\perp}^{2}\over Q^{2}}
	\ +\ \Bigl[(\hat t,a)\leftrightarrow(\hat u,b)\Bigr]
	\ ,\label{eq:pp}\nonumber\\
	B_{2(++)}
	&=&-\,{{\hat s}^{2}p_{\perp}^{2}Q^{2}\over2{\hat u}{\hat t}}
	\left({\hat s}^{2}+Q^{4}\right)\ln{Q_{\perp}^{2}\over Q^{2}}	\nonumber\\
	&&+{{\hat s}p_{\perp}^{2}(Q^{2}-{\hat u})^{3}\over2{\hat u}
	{\hat t}(Q^{2}-{\hat t})}
	\left((Q^{2}-{\hat t}){\hat 
	s}+Q^{2}{\hat u}\right)\left[{2\over3}+{Q^{2}\over Q_{\perp}^{2}}
	-{10\over3}
	{Q^{6}\over Q_{\perp}^{6}}\right]\nonumber\\
	&&-{{\hat s}p_{\perp}^{2}(Q^{2}-{\hat u})^{2}\over2{\hat u}
	{\hat t}(Q^{2}-{\hat t})^{2}}
	\Biggl[3(Q^{2}-{\hat t})^{3}Q^{2}
	+(Q^{2}-{\hat t})Q^{2}(2{\hat u}{\hat t}+m_{H}^{4})\nonumber\\
	&&\qquad\qquad+(Q^{2}-{\hat t})^{2}\left({\hat 
	s}^{2}+4m_{H}^{2}Q^{2}-{\hat u}(Q^{2}+m_{H}^{2})\right)\nonumber\\
	&&\qquad\qquad
	-{\hat u}^{2}Q^{4}+{\hat u}{\hat t}^{2}m_{H}^{2}
	\Biggr]\left[
	1	-2{Q^{4}\over Q_{\perp}^{4}}\right]\\
	&&+{{\hat s}p_{\perp}^{2}(Q^{2}-{\hat u})\over2{\hat u}
	{\hat t}(Q^{2}-{\hat t})}
	\Biggl[3Q^{2}{\hat s}(Q^{2}+{\hat s})(Q^{2}-{\hat t})
	-{\hat t}{\hat s}^{3}\nonumber\\
	&&\qquad\qquad+m_{H}^{2}Q^{2}{\hat s}^{2}
	+Q^{2}{\hat u}(m_{H}^{2}-Q^{2})^{2}
	\Biggr]\left[
	1-2{Q^{2}\over Q_{\perp}^{2}}\right]\nonumber\\
	&&+{{\hat s}p_{\perp}^{2}\over12{\hat u}{\hat t}}\biggl(
	-2{\hat s}^{4}+6{\hat s}m_{H}^{2}{\hat t}({\hat t}-m_{H}^{2})
	+2{\hat s}m_{H}^{6}+8Q^{2}{\hat s}({\hat s}-Q^{2})^2
	-2{\hat u}{\hat t}{\hat s}Q^{2} \nonumber\\
	&&\qquad\qquad
	+7{\hat s}^2m_{H}^{2}Q^{2}
	-2{\hat s}Q^{4}m_{H}^{2}
     -m_{H}^{6}Q^{2} +3m_{H}^{2}Q^{6} -4{\hat u}{\hat t}m_{H}^{2}Q^{2}
	 \biggr)  \nonumber\\
	&&+{11\over6}{{\hat s}^{3}p_{\perp}^{2}Q^{4}\over{\hat u}{\hat t}}
	-{{\hat s}^{2}p_{\perp}^{2}m_{H}^{4}Q^{2}\over3{\hat t}^{2}}
	\ +\ \Bigl[(\hat t,a)\leftrightarrow(\hat u,b)\Bigr]\ .\nonumber
\end{eqnarray}

\subsection{$gq\rightarrow H+X$ terms}

We write the gluon-quark ``nonsingular'' contribution as
\begin{eqnarray}
G^{({\rm 2R, ns})}_{gq}&=& {1\over\hat s^2p_\perp^2Q^2}\Biggl\{C_F^2\Bigl(C_{1(+-)}
+C_{1(-+)}+C_{1(++)}+C_{1(--)}
\Bigr)\\
&&\qquad\,+N_cC_F\Bigl(C_{2(+-)}+C_{2(-+)} +C_{2(++)} 
+C_{2(--)}+C_{2\epsilon}\Bigr)\Biggr\}\nonumber\ ,
\end{eqnarray}
where the terms, $C_{i(\pm\pm)}$, come from the $Hggq\bar q$ 
squared amplitudes with helicity configurations  $(q-,\bar q+,g_1\pm, g_2\pm)$,  plus the 
helicity-flipped configuration that contributes identically.  To be precise, the helicities
are written for all final-state momenta, before crossing the particles $g_1$ and $\bar q$ into
the initial-state to obtain the scattering $g_1q\rightarrow g_2qH$.  The term $C_{2\epsilon}$ comes from the ${\cal O}(\epsilon)$  corrections to the squared 
$Hggq\bar q$ matrix elements that
have not been included
in $G^{({\rm 2s})}_{gq}$, eq.~(\ref{eq:gqsing}).  
Of course, $G^{({\rm 2R, ns})}_{qg}$ is obtained from these expressions by $(\hat t,a)\leftrightarrow(\hat u,b)$.

We obtain:
\begin{eqnarray}
	C_{1(+-)}
	& = &   
	-2{\hat s}^2p_{\perp}^{2}Q^2\ln{p_{\perp}^{2}/Q_{\perp}^{2}}\ ,\\
	C_{2(+-)}&=&0
		\label{eq:pmqqb}\ ,\\
	C_{1(-+)}
	& = &    
	\hat s^2p_\perp^2Q^2-{3\hat s^2p_\perp^2\hat t^2\over2\hat u}
	\nonumber\\
	&&
	+{p_\perp^2\over2Q_\perp^2}\left(
      	{\hat s}(Q^2-\hat t)^3+Q^2(Q^2-\hat u)^3\right)
	\left[-3+10{Q^2\over Q_\perp^2}-6{Q^4\over Q_\perp^4}\right]
	\ ,\\
	C_{2(-+)}
	&=&	 
	{p_\perp^2Q^2\over Q_\perp^4}\left(
      	{\hat s}(Q^2-\hat t)^3+Q^2(Q^2-\hat u)^3\right)
	\left(-2+3{Q^2\over Q_\perp^2}\right)\nonumber\\
	&&	+2\hat s^2p_\perp^2Q^2
	+4\hat s^2p_\perp^2Q^2\ln{p_{\perp}^{2}/Q_{\perp}^{2}}\ ,\\
	C_{1(++)}
	& = &    	
	-{3\hat s^4p_\perp^2\over2\hat u}-{\hat s p_\perp^2 Q^2A^2\over{\hat u}}L_{2b}
	\nonumber\\
	&&+{\hat s p_\perp^2\over{\hat u}{\hat t}^2}L_{1a}
	\biggl[(A-m_H^2)^2\hat s\hat t^2-2Q^2m_H^2\hat u\hat t A
	-Q^2m_H^4(Q^2-\hat t)\hat u\biggr]\nonumber\\
	&&+{\hat s p_\perp^2\over{\hat u}B}L_{3}
	\biggl[(\hat s+Q^2-m_H^2)\Bigl(\hat s(A-m_H^2)^2+Q^2A^2\Bigr)
	-4\hat s Q^2A(A-m_H^2)\biggr]\nonumber\\
	&&+{1\over2}\hat s p_\perp^2(Q^2-\hat t)^2
	\left({\hat s\over(Q^2-\hat u)}-{Q^2\over\hat u}\right)
	\left[-3+10{Q^2\over Q_\perp^2}-6{Q^4\over Q_\perp^4}\right]
	\nonumber\\
	&&+{1\over2}\hat s p_\perp^2(Q^2-\hat u)^2
	\left({Q^2\over(Q^2-\hat t)}+{\hat s\over\hat u}\right)
	\left[-3+10{Q^2\over Q_\perp^2}-6{Q^4\over Q_\perp^4}\right]
	\nonumber\\
	&&+{\hat s p_\perp^2(Q^2-\hat t)\over\hat u(Q^2-\hat u)}
	\biggl(2\hat s\hat u(\hat s+\hat t)-Q^2(4m_H^2Q^2-Q^2\hat t-m_H^2\hat u\nonumber\\
	&&\qquad\qquad\qquad\qquad\qquad\qquad\qquad
	-2\hat u\hat t)\biggr)
	\left[-1+2{Q^2\over Q_\perp^2}\right]\nonumber\\
	&&+{\hat s p_\perp^2(Q^2-\hat u)\over\hat u(Q^2-\hat t)}
	\biggl(\hat t\hat s^2-2\hat u\hat t\hat s+2Q^2\hat u(Q^2-\hat t)\biggr)
	\left[-1+2{Q^2\over Q_\perp^2}\right]\nonumber\\
	&&+{\hat s p_\perp^2Q^2m_H^2(\hat t+\hat u)\over\hat t}
	-{2\hat s p_\perp^2Q^4m_H^4\over\hat t^2}
	+{\hat s^2 p_\perp^2Q^2m_H^4\over2\hat u^2}\nonumber\\
	&&+{\hat s p_\perp^2\over2\hat u}
	\biggl[-2(Q^2+m_H^2)\hat s\hat u+2m_H^2\hat s^2+m_H^4\hat s\nonumber\\
	&&\qquad\quad
	+m_H^2Q^2\Bigl(2(\hat s-Q^2)+3m_H^2-\hat u\Bigr)+5Q^2\hat s(\hat s-Q^2)\biggr]
	\ ,\\
	C_{2(++)}
	&=&	     
{1\over2}{\hat s p_\perp^2A^2 (1-z_a)}(L_{2a}-L_{1a})
	+{\hat s p_\perp^2(m_H^2-\hat t)A^2(1-z_b)\over2\hat u}(L_{1b}-L_{2b})\nonumber\\
	&&+{\hat s p_\perp^2(\hat s-Q^2)^3\over2\hat u}(L_{1b}-L_{1a})
	+{\hat s p_\perp^2Q^2A^2\over(m_H^2-\hat u)}(L_{1b}+L_{2a})\nonumber\\
	&&+{\hat s p_\perp^2Q^2\over\hat u^2}L_{1b}
	\biggl[2\hat u(\hat s-Q^2)^2+4m_H^2(m_H^2-\hat t)A
	-2m_H^4(Q^2-\hat t)-m_H^6\biggr]\nonumber\\
	&&-{1\over2}\hat s p_\perp^2(Q^2-\hat t)^2
	\left({\hat s\over(Q^2-\hat u)}-{Q^2\over\hat u}\right)
	\left[-3+10{Q^2\over Q_\perp^2}-6{Q^4\over Q_\perp^4}\right]
	\nonumber\\
	&&-{1\over2}\hat s p_\perp^2(Q^2-\hat u)^2
	\left({Q^2\over(Q^2-\hat t)}+{\hat s\over\hat u}\right)
	\left[-3+10{Q^2\over Q_\perp^2}-6{Q^4\over Q_\perp^4}\right]
	\nonumber\\
	&&+{\hat s p_\perp^2(Q^2-\hat t)\over2\hat u(Q^2-\hat u)}
	\biggl((-3\hat sp_\perp^2+\hat u^2-Q^4)(\hat s+Q^2)
	-m_H^2\hat s\hat u\nonumber\\
	&&\qquad\qquad\qquad+2Q^4(Q^2-\hat u)\biggr)
	\left[-1+2{Q^2\over Q_\perp^2}\right]\nonumber\\
	&&+{\hat s p_\perp^2(Q^2-\hat u)\over2\hat u(Q^2-\hat t)}
	\biggl(3\hat sp_\perp^2(\hat s+Q^2)-\hat uQ^2(Q^2-\hat t)
	+3Q^2m_H^2(Q^2-\hat u)\nonumber\\
	&&\qquad\qquad\qquad+\hat s\hat t(\hat u+\hat s)\biggr)
	\left[-1+2{Q^2\over Q_\perp^2}\right]\nonumber\\
	&&+{\hat s p_\perp^2m_H^2Q^2\over2\hat u^2}
	\biggl[2(\hat s-Q^2)^2-2m_H^2(\hat s-m_H^2)-\hat u(Q^2-\hat 	u)-4m_H^2Q^2\biggr]\nonumber\\	
	&&+{\hat s^2 p_\perp^2(\hat u-\hat t)(m_H^2+Q^2)\over2\hat u}	-{8(\hat s p_\perp^2m_H^2Q^2)^2\over\hat u^4}-{2(\hat s p_\perp^2m_H^2Q^2)^2\over\hat u^4}L_{1b}\ ,\\
	C_{1(--)}
	& = &    
	{\hat s^2p_\perp^2\hat t^2\over{\hat u}}L_{1a}
-{\hat s p_\perp^2Q^2(m_H^2-\hat t)^2\over{\hat u}}L_{2b}
	+{\hat s p_\perp^2m_H^2Q^2\over B^2}\left(\hat t(\hat u+\hat t)-2m_H^2Q^2\right)
	\nonumber\\
	&&+{\hat s p_\perp^2\over{\hat u}B}\Biggl\{
	\hat t^2B^2-m_H^2\hat t^2(\hat u+\hat t)+2Q^4m_H^4
        	+Q^2m_H^4(3\hat t-\hat u)\nonumber\\
	&&\qquad\quad+{Q^2m_H^4\hat u\over B^2}\left(-\hat t(\hat u+\hat t)+2m_H^2Q^2
	+Q^2(\hat t-\hat u)\right)\Biggr\}L_3\ ,\\
	C_{2(--)}
	&=&	     
{\hat s p_\perp^2\hat t^2 Q^2\over2\hat u}(L_{1a}+3L_{1b})
	+{1\over2}\hat s p_\perp^2\hat t^2(1-z_a)(L_{2a}-L_{1a})\nonumber\\
	&&+{\hat s p_\perp^2Q^2(m_H^2-\hat t)^3z_b\over2\hat u^2}(L_{2b}-L_{1b})
	+{\hat s p_\perp^2\hat t^2Q^2\over(m_H^2-\hat u)}(L_{1b}+L_{2a})\nonumber\\
	&&+{\hat s^2 p_\perp^2\hat t^2\over2\hat u}(L_{1b}-L_{1a})
	+{\hat s p_\perp^2m_H^2Q^2\over\hat u^2}
	\left(4\hat t(\hat t-m_H^2)+m_H^4\right)L_{1b}\nonumber\\
	&&-{2(\hat s p_\perp^2m_H^2Q^2)^2\over\hat u^4}L_{1b}
	+{\hat s p_\perp^2\hat t^2m_H^2Q^2\over\hat u^2}
	-{8(\hat s p_\perp^2m_H^2Q^2)^2\over\hat u^4}\ ,\\
C_{2\epsilon}&=&
{4(\hat sp_\perp^2m_H^2Q^2)^2\over\hat u^4}\ .
\end{eqnarray}

\subsection{$q_i\bar q_i\rightarrow H+X$ terms}

We write the quark-antiquark (same flavor) ``nonsingular'' contribution as
\begin{eqnarray}
G^{({\rm 2R, ns})}_{q_i\bar q_i}&=& {1\over\hat s^2p_\perp^2Q^2}\Biggl\{2C_F^3\Bigl(D_{1(+-)}
+D_{1(++)}
\Bigr)+2N_cC_F^2\Bigl(D_{2(+-)}+D_{2(++)} 
\Bigr)\nonumber\\
&&\qquad\,+n_fC_F^2E_1
 +C_F^2E_2
+{C_F^2\over N_c}E_3\Biggr\}\ ,
\end{eqnarray}
where the terms, $D_{i(+-)}$ and $D_{i(++)}$, come from the $Hggq\bar q$ squared amplitudes with unlike-helicity gluons and like-helicity gluons, respectively.   
The $E_1$ terms arise from the $s$-channel diagrams, the
$E_2$ terms come from the $t$- and $u$-channel diagrams, and the $E_3$ terms come from
the cross terms.  In the case of quark-antiquark and quark-quark scattering contributions,
all ${\cal O}(\epsilon)$  corrections have been included in the formulae in the main text
for $G^{({\rm 2s})}_{q\bar q}$, eqs.~(\ref{eq:qiaising}) and (\ref{eq:qiajsing}).

We obtain from the $Hggq\bar q$ squared matrix elements:
\begin{eqnarray}
	D_{1(+-)}
	& = & 
        	-{\hat s}^{2}p_{\perp}^{2}Q^{2}
	\left(1+\ln{p_{\perp}^{2}/Q_{\perp}^{2}}\right)
	-{\hat s^3p_\perp^4z_a(1-z_a)\over\hat t}-\hat s^3p_\perp^2(1-z_a)^2
	\nonumber\\
	&&+\ \Biggl[{\hat t}\leftrightarrow{\hat u}\Biggr]
	\ ,\label{eq:qqpm}\\
	D_{2(+-)}
	&=&-{1\over3}{\hat s}^{2}p_{\perp}^{2}Q^{2}
	-{\hat sp_\perp^2\hat t^2\over6 z_a^{2}}
	\left(11-12{Q^2\over Q_\perp^2}+3{Q^4\over Q_\perp^4}\right)	+{11\hat sp_\perp^2\hat t^2\over6 }\nonumber\\
	&&+\ \Biggl[{\hat t}\leftrightarrow{\hat u}\Biggr]
	\ ,\\
	D_{1(++)}
	& = & {\hat s^2p_\perp^2\hat u^2(1-z_b)\over A}(L_{2b}-L_{1b})
	+{\hat s^2p_\perp^2(m_H^2-\hat u)^2(1-z_a)\over A}(L_{2a}-L_{1a})\nonumber\\
	&&+{\hat sp_\perp^2m_H^2Q^2(\hat sp_\perp^2+\hat u\hat t)\over \hat t^2}L_{1a}
	-{2\hat s^2p_\perp^2m_H^4Q^2\over AB}L_3	
	+{{\hat s}p_{\perp}^{2}m_H^2Q^{2}(2\hat sp_\perp^2-\hat u\hat t)
	\over\hat t^2}\nonumber\\
	&&+\ \Biggl[{\hat t}\leftrightarrow{\hat u}\Biggr]
	\ ,\label{eq:qqpp}\\
	D_{2(++)}
	&=&{\hat sp_\perp^2\hat u^2(m_H^2-\hat t)(1-z_b)\over 2A}(L_{1b}-L_{2b})
	+{\hat sp_\perp^2(m_H^2-\hat u)^3(1-z_a)\over 2A}(L_{1a}-L_{2a})\nonumber\\
	&&-{1\over2}\hat sp_\perp^2\hat u^2(L_{1a}+L_{1b})
	+{6(\hat sp_\perp^2m_H^2Q^2)^2\over B^4}
	-{\hat sp_\perp^2m_H^2Q^2\hat u^2\over B^2}\nonumber\\
	 &  & + L_{3} \left\{
	 {{\hat s}p_{\perp}^{2}{\hat u}^{2}({\hat u}+{\hat t})
	 \over B}+\left({\hat s}p_{\perp}^{2}m_{H}^{2}Q^{2}\right)^{2}
	 \left[{1\over AB^{3}}-{3A\over B^{5}}\right]\right.\nonumber\\
	 &  &\left.\qquad +{\hat s}p_{\perp}^{2}m_{H}^{2}Q^{2}\left[
	 {(\hat t-3\hat u)\over 2B}+
	 {A(B^2+2\hat u^2)\over4B^{3}}+{(
	 {\hat t}^{2}-6{\hat t}{\hat u}+7{\hat u}^{2}
	 )\over4AB}\right]\right\}\nonumber\\
	&&+\ \Biggl[{\hat t}\leftrightarrow{\hat u}\Biggr]
	\ .
\end{eqnarray}

We obtain from the $Hq\bar qq\bar q$ squared matrix elements:
\begin{eqnarray}
E_1&=&{8\over3}\hat s^2p_\perp^2Q^2-{4\over3}\hat sp_\perp^2 m_H^2Q^2\ ,\\
E_2&=&
{\hat sp_\perp^2Q^2(Q^2-p_\perp^2)\over Q_\perp^4}\left((\hat t/z_a)^2
+(\hat u/z_b)^2\right)+2\hat s^2p_\perp^2Q^2\ ,\label{eq:eii}\\
E_3&=&-2\hat sp_\perp^2\left(\left(\hat u+\hat t-2Q^2\right)^2-2\hat sp_\perp^2\right)\ln{p_{\perp}^{2}/Q_{\perp}^{2}}
\nonumber\\
&&-{\hat sp_\perp^2Q^2(2Q_\perp^2+Q^2)\over Q_\perp^4}\left((\hat t/z_a)^2
+(\hat u/z_b)^2\right)\,-6\hat s^2p_\perp^2Q^2\ .
\label{eq:eiie}\end{eqnarray}

\subsection{$q_i \bar q_j\rightarrow H+X$ terms}

The quark-antiquark (different flavors) ``nonsingular'' contribution comes from $Hq_i\bar q_iq_j\bar q_j$ squared
amplitudes and can be written
\begin{eqnarray}
G^{({\rm 2R, ns})}_{q_i\bar q_j}&=& {1\over\hat s^2p_\perp^2Q^2}\Biggl\{C_F^2E_2
\Biggr\}\nonumber\ ,\label{qqdif}
\end{eqnarray}
where $E_2$ was given in eq.~(\ref{eq:eii}).
The quark-quark (different flavors) ``nonsingular" contribution is also given by this
same expression, eq.~(\ref{qqdif}).

\subsection{$q_i q_i\rightarrow H+X$ terms}

The quark-quark (same flavor) ``nonsingular'' contribution comes from $Hq_i\bar q_iq_i\bar q_i$
squared amplitudes and can be written
\begin{eqnarray}
G^{({\rm 2R, ns})}_{q_iq_i}&=& {1\over\hat s^2p_\perp^2Q^2}\Biggl\{C_F^2E_2
+{C_F^2\over N_c}E_4\Biggr\}\nonumber\ ,
\end{eqnarray}
where $E_2$ was given in eq.~(\ref{eq:eii}),
and the subleading term in $1/N_c$ is
\begin{eqnarray}
E_4&=&{2\hat sp_\perp^2Q^2(\hat s^2+Q^4)\over Q_\perp^2}\ln{p_{\perp}^{2}/Q^{2}}
+4\hat s^2p_\perp^2Q^2\ln{p_{\perp}^{2}/Q_\perp^{2}}\ .
\end{eqnarray}


\section{Phase space for integrals over momentum fractions}
\label{app:phasespace}

After integration over the angular variables of the two emitted
partons, we are left with the integrals over $0<x_a,x_b<1$ with
the restriction to $Q^2>0$.  A parametrization of this
phase space was given in Ref.~\cite{EMP}.  Using a change of 
variables, we can write the phase space as a sum of two double integrals,
where the inner integral is over the momentum fraction $z_a$ or
$z_b$, defined in eq.~(\ref{eq:zb}).  Specifically, for the case
where the inner integral is over $z_a$, we write
\begin{eqnarray}
	\int_{0}^{1}dx_{a}\int_{0}^{1}dx_{b}\,\Theta(Q^{2})F(x_{a},x_{b})
	&=& \left\{\int_{x_{b}^{0}}^{1-\delta}{dz_{b}^\prime\over 
	z_{b}^\prime}
	\int_{x_{a}^{0}(1+\lambda_{b})}^{1}{dz_{a}\over z_a}
	\left[{m_\perp^2(1+\lambda_b)\over z_{a}z_b^\prime}\right]\right.
	\label{eq:phasethree}\\
	&&\left.+
	\int_{x_{a}^{0}}^{1-\delta}{dz_{a}^\prime\over 
	z_{a}^\prime}
	\int_{x_{a}^{0}/z_{a}^\prime}^{1}{dz_{a}\over z_a}
	\left[{p_\perp^2\over (1-z_a^\prime)^2z_{a}}\right]	
	\right\}
	{F(x_{a},x_{b})\over s}\ ,
    \nonumber
\end{eqnarray}
with
\begin{eqnarray}
\delta&=&p_{\perp}/(m_{\perp}+p_{\perp})\nonumber\\
	x_{a}^{0} & = & {m_{\perp}e^{y_{H}}\over\sqrt{s}}
	\nonumber  \\
	x_{b}^{0} & = & {m_{\perp}e^{-y_{H}}\over\sqrt{s}}\ .
	\label{eq:xnaughts}
\end{eqnarray}
In the first double integral
\begin{eqnarray}
	x_{a} & = & {x_{a}^{0}\over z_{a}}(1+\lambda_{b})\nonumber\\
	x_{b} & = & {x_{b}^{0}\over z^{\prime}_{b}}
	\nonumber  \\
	{\hat s} &=& {m_{\perp}^{2}\over z_{a}z^{\prime}_{b}}
	(1+\lambda_{b})\nonumber\\
	{\hat t} & = & - {m_{\perp}^{2}\over z_{b}^\prime}(1-z_b^\prime)
	(1+\lambda_{b})\nonumber\\
	{\hat u} & = & m_H^2-{m_\perp^2\over z_a}(1+\lambda_{b})
	\nonumber\\
	Q^{2} & =& {m_{\perp}^{2}\over z_az_{b}^\prime}(1-z_a)(1-z_b^\prime)
	(1+\lambda_{b})\ ,\label{eq:i1var}
\end{eqnarray}
where 
\begin{equation}
	\lambda_{i}\ =\ {p_{\perp}^{2}\over m_{\perp}^{2}}
	{z_{i}^\prime\over(1- z_{i}^\prime)}\ .
	\label{eq:lambdaa}
\end{equation}
In the second double integral
\begin{eqnarray}
	x_{a} & = & {x_{a}^{0}\over z_{a}z_{a}^{\prime}}\nonumber\\
	x_{b} & = & x_{b}^{0}(1+\lambda_{a})
	\nonumber  \\
	{\hat s} &=& {m_{\perp}^{2}\over z_{a}z^{\prime}_{a}}
	(1+\lambda_{a})\nonumber\\
	{\hat t} & = & -{p_{\perp}^{2}\over1-z_{a}^\prime}\nonumber\\
	{\hat u} & = & m_{H}^{2}
	-{m_{\perp}^{2}\over z_az_{a}^{\prime}}
	 \nonumber\\
	Q^{2} & =& {p_{\perp}^{2}(1-z_a)\over z_a(1-z_a^\prime)}
	\ .
\end{eqnarray}
For very large $|y_H|$ the limits on the integrals may be further restricted
by the requirements $x_a,x_b<1$.

The two double integrals become particularly simple 
in the limit $p_\perp/m_\perp\rightarrow0$, and they have 
simple physical interpretations.
The first double integral becomes two independent convolutions,
corresponding to independent emissions 
off the $a$ and the $b$ partons. The second double
integral becomes a double convolution,
corresponding to two successive emissions off the $a$ parton.  
We can alternatively write the phase space with the inner
integration over $z_{b}$ by replacing $a\leftrightarrow b,\ {\hat 
t}\leftrightarrow{\hat u}$ everywhere.  
The choice of the inner integral over
$z_{a}$ or $z_{b}$ is a matter of convenience.  We have found
it most natural (and efficient in the small-$p_\perp$ limit)
to compute the ``singular'' terms that are
written specifically as functions or distributions in
$z_{a(b)}$ in section~\ref{sec:total} using
the phase space with the inner integral over $z_{a(b)}$.
The other terms are computed with the phase space symmetrized
over $a\leftrightarrow b$.  


\section{Details of the small-$p_\perp$ limit}
\label{app:smallpt}

As discussed in section~\ref{sec:smallpt}, the resummed formula,
eq.~(\ref{eq:resumm}), can be expanded as a
power series in $\alpha_s$, yielding 
\begin{equation}
	{d\sigma\over dp_{\perp}^{2}dy_{H}}\Biggr|_{p_{\perp}\ll m_H}
\ =\
	{\sigma_{0}\over s}{m_{H}^{2}\over p_{\perp}^{2}}
\left[
\sum_{m=1}^2\sum_{n=0}^{2m-1}\left(
	{\alpha_{s}\over2\pi}\right)^{m}
	\,_{m}C_{n}\,\left(\ln{m_{H}^{2}\over 
	 p_{\perp}^{2}}\right)^{n}
+{\cal O}(\alpha_s^3)\right]
\ ,\label{eq:smallpt2}
\end{equation}
where the coefficients are
\begin{eqnarray}
	_{1}C_{1}&=& A^{(1)}f_{g}({\bar x}_{a}^{0})f_{g}({\bar x}_{b}^{0})\,,
	\nonumber\\
	_{1}C_{0}&=& B^{(1)}f_{g}({\bar x}_{a}^{0})f_{g}({\bar x}_{b}^{0})
	+\Bigl(P_{gi}\circ f_{i}\Bigr)({\bar x}_{a}^{0})f_{g}({\bar x}_{b}^{0})
	+f_{g}({\bar x}_{a}^{0})\Bigl(P_{gi}\circ f_{i}\Bigr)({\bar x}_{b}^{0})
\ ,\nonumber\\
	_{2}C_{3}&=&
-{1\over2}\left[A^{(1)}\right]^2f_g({\bar x}_a^0)f_g({\bar x}_b^0)\ ,\nonumber
\\
	_{2}C_{2}&=&-{3\over2}A^{(1)}\Biggl[
\Bigl(P_{gi}\circ f_{i}\Bigr)({\bar x}_{a}^{0})f_{g}({\bar x}_{b}^{0})+
f_{g}({\bar x}_{a}^{0})\Bigl(P_{gi}\circ f_{i}\Bigr)({\bar x}_{b}^{0})\Biggr]
\nonumber\\
&&+A^{(1)}\left[\beta_0-{3\over2}B^{(1)}\right]
f_g({\bar x}_a^0)f_g({\bar x}_b^0)\ ,\nonumber\\
	_{2}C_{1}& =& 
\left[\beta_0-2B^{(1)}-A^{(1)}\ln{\mu_F^2\over m_H^2}\right]\Bigl(P_{gi}\circ f_{i}\Bigr)({\bar x}_{a}^{0})f_{g}({\bar x}_{b}^{0})
+A^{(1)}\Bigl(C^{(1)}_{gi}\circ f_i\Bigr)({\bar x_a^0})f_{g}({\bar x}_{b}^{0})\nonumber\\
&&
-\Bigl(P_{gi}\circ f_{i}\Bigr)({\bar x}_{a}^{0})
\Bigl(P_{gj}\circ f_{j}\Bigr)({\bar x}_{b}^{0})	
-\Bigl(P_{gi}\circ P_{ij}\circ f_{j}\Bigr)({\bar x}_{a}^{0})f_{g}({\bar x}_{b}^{0})\nonumber\\
&&-{1\over2}\left[\left[B^{(1)}\right]^2-A^{(2)}-\beta_0B^{(1)}-3\beta_0A^{(1)}\ln{\mu_R^2\over m_H^2}\right]
f_g({\bar x}_a^0)f_g({\bar x}_b^0)\nonumber\\
&&+\{a\leftrightarrow b\}\ ,\nonumber\\
	_{2}C_{0}& =& -
\Biggl[\Bigl(P_{gi}\circ f_{i}\Bigr)({\bar x}_{a}^{0})
\Bigl(P_{gj}\circ f_{j}\Bigr)({\bar x}_{b}^{0})
 +\Bigl(P_{gi}\circ P_{ij}\circ f_{j}\Bigr)({\bar x}_{a}^{0})
f_{g}({\bar x}_{b}^{0})
\Biggr]
\ln{\mu_F^2\over m_H^2}\nonumber\\
&& +\Biggl[
3\beta_0\ln{\mu_R^2\over m_H^2}
-B^{(1)}\ln{\mu_F^2\over m_H^2}\Biggr]
\Bigl(P_{gi}\circ f_{i}\Bigr)({\bar x}_{a}^{0})
f_{g}({\bar x}_{b}^{0})\nonumber\\
&&
+\Bigl(C^{(1)}_{gi}\circ f_{i}\Bigr)({\bar x}_{a}^{0})
\Bigl(P_{gj}\circ f_{j}\Bigr)({\bar x}_{b}^{0})
+\Bigl(C^{(1)}_{gi}\circ P_{ij}\circ f_{j}\Bigr)({\bar x}_{a}^{0})
f_{g}({\bar x}_{b}^{0})
\nonumber\\
&&+\left[\zeta(3)\left[A^{(1)}\right]^2+{1\over2}B^{(2)}
+{3\over2}\beta_0B^{(1)}\ln{\mu_R^2\over m_H^2}\right]
f_g({\bar x}_a^0)f_g({\bar x}_b^0)\nonumber\\
&&+\left(B^{(1)}-\beta_0\right)\Bigl(C^{(1)}_{gi}\circ f_{i}\Bigr)({\bar x}_{a}^{0})
f_{g}({\bar x}_{b}^{0})
+\Bigl(P^{(2)}_{gi}\circ f_{i}\Bigr)({\bar x}_{a}^{0})
f_{g}({\bar x}_{b}^{0})\nonumber\\
&&+\{a\leftrightarrow b\}
	\ .
	\label{eq:mCn}
\end{eqnarray}
In these equations the indices $i$ and $j$ 
are implicitly summed over all massless partons
($g,q_f,\bar q_f$), and the parton density functions are
always evaluated at the factorization scale $\mu_F$; that is
$f_i(x)\equiv f_i(x,\mu_F)$.  The parameters, $A^{(1,2)}$, $B^{(1,2)}$,
and the functions $C^{(1)}_{ij}(z)$  were given in eq.~(\ref{eq:params}).
The function $P^{(2)}(z)$ is the NLO splitting function.
The expressions for the $_mC_n$ are precisely the same as those 
given in Ref.~\cite{ak}, except for the additional factor of 3 in front of
each occurrence of $\ln(\mu_R^2/m_H^2)$.  This factor, of course,
is due to the fact that the LO $p_\perp$ spectrum for $gg\rightarrow
H+X$ with $p_\perp>0$ begins at ${\cal O}(\alpha_s^3)$.

We have explicitly checked that our NLO calculation agrees 
with this asymptotic expression in the limit 
$p_\perp\ll m_H$.  We give 
some of the details of how this check was performed here.

The coefficients in eq.~(\ref{eq:mCn}) can be separated into
contributions:
\begin{equation}
_mC_n\ =\ _mC_n^{(gg)}+\,_mC_n^{(gq)}+\,_mC_n^{(qg)}+\,_mC_n^{(qq)}\ ,
\end{equation}
where the superscripts label which type of partons come 
from the $a$ and $b$ hadrons, and the label $q$ actually 
implies a sum over all light quarks and antiquarks.   The 
quark-quark contributions, $_mC_n^{(qq)}$, are particularly
simple, so we begin the discussion with them.

First we note that $g_{q\bar q}$ is not singular as $p_\perp\rightarrow0$,
so that the LO cross section does not contribute to the small $p_\perp$
limit; {\it i.e.}, $_1C_n^{(qq)}=0$.  Similarly, all terms at NLO
which are proportional to $\delta(Q^2)$ do not contribute.  In fact,
it is easy to see that the 
``singular'' terms that contribute are independent of quark
flavors and of whether it is quark-quark or quark-antiquark scattering.
Finally, we note that none of the ``nonsingular'' quark-quark
terms contribute in the small-$p_\perp$ limit.  The general argument
is as follows.  Since the
$G^{({\rm 2R, ns})}_{ij}$ terms are defined to have no explicit 
singularities as $p_\perp\rightarrow0$, they can only contribute
at $1/p_\perp^2$ if they obtain the singularity 
through integration over the phase
space near $Q^2=0$.  Effectively, this implies that ``nonsingular''
terms can only contribute if they are ${\cal O}(1/p_\perp^4)$ when
$Q^2$ is treated as ${\cal O}(p_\perp^2)$ and either $\hat u$ or
$\hat t$ is treated ${\cal O}(p_\perp^2)$.

Thus we get
\begin{eqnarray}
G^{(2)}_{qq}\Bigr|_{p_\perp\ll m_H} 
 &\approx&\Biggl\{	\left({1\over -\hat t}\right)\Biggl[-
P_{gq}(z_a)\ln{\mu_F^2\over Q^2}+C^{\epsilon}_{gq}(z_a)\Biggr]
g_{g q,a}(z_a)\nonumber\\
&&\quad
+{C_F^2}\left[{ (\hat s-Q^2)^2+\left(\hat u+\hat t-2Q^2\right)^2
\over\hat s}
\right]\,{1\over p_\perp^2}\ln {p_\perp^2\over Q_\perp^2}\nonumber\\
&&
+\Bigl[(\hat t,a)\leftrightarrow(\hat u,b)\Bigr]\Biggr\}
\ .\label{eq:qqsmallpt}
\end{eqnarray}
Inserting this into eqs.~(\ref{eq:series}) and (\ref{eq:master}), this is
easily integrated using the phase space parametrization of 
eq.~(\ref{eq:phasethree}).  Since there are no singular terms in the
integrands as $z^\prime_{a,b}\rightarrow1$ or $z_{a,b}\rightarrow1$, 
the $p_\perp\ll m_H$ limit
is trivially taken, with only the first integral of eq.~(\ref{eq:phasethree})
contributing.  The result is
\begin{eqnarray}
	_{2}C_{3}^{(qq)}&=& _{2}C_{2}^{(qq)}\ =\ 0\ ,\nonumber\\
	_{2}C_{1}^{(qq)}& =& -\Bigl(P_{gq}\circ f_{q}\Bigr)({\bar x}_{a}^{0})
\Bigl(P_{gq}\circ f_{q}\Bigr)({\bar x}_{b}^{0})	
+\{a\leftrightarrow b\}\ ,\nonumber\\
	_{2}C_{0}^{(qq)}& =& -\Bigl(P_{gq}\circ f_{q}\Bigr)({\bar x}_{a}^{0})
\Bigl(P_{gq}\circ f_{q}\Bigr)({\bar x}_{b}^{0})\ln{\mu_F^2\over m_H^2}
+\Bigl(C^{(1)}_{gq}\circ f_{q}\Bigr)({\bar x}_{a}^{0})
\Bigl(P_{gq}\circ f_{q}\Bigr)({\bar x}_{b}^{0})\nonumber\\
&&+\{a\leftrightarrow b\}
	\ .
\label{eq:qq20}
\end{eqnarray}

The gluon-quark contributions are obtained similarly, 
with two added complications.  The first is that the ``nonsingular''
terms now contribute to the cross section.  The simplest way to handle
these terms is to note that they are singular as 
$\hat u\sim {\cal O}(p_\perp^2)\rightarrow0$
but not as $\hat t\sim {\cal O}(p_\perp^2)\rightarrow0$.  
Thus, we can use the phase space of eq.~(\ref{eq:phasethree})
with the replacement $a\leftrightarrow b$, and only the
second integral contributes in the small-$p_\perp$ limit.  Since there
are no subtleties in this integral, one can directly take the limit
$p_\perp/m_H\rightarrow0$ everywhere, leaving a double convolution,
which can easily be converted to a single convolution.  

The second complication is that the integrands have singular behavior as
both $z^\prime_{a,b}\rightarrow1$ and as $z_{a,b}\rightarrow1$.  The
first of these can be handled by the following identity as $p_\perp/m_H
\rightarrow0$:
\begin{eqnarray}
	\int_{x_{i}^{0}}^{1-\delta}dz^\prime_{i}\,F(z^\prime_{i})\,{\ln^k(1-z^\prime_i)\over1-z^\prime_i}
	\quad & \longrightarrow &\quad\quad
	\int_{x_{i}^{0}}^{1}dz^\prime_{i}\,F(z^\prime_{i})\, 
	 \left[{\ln^{k}{(1-z^\prime_{i})}\over1-z^\prime_{i}}\right]_{+}\nonumber\\
&&\quad	 -F(1)\,{1\over k+1}\left(-{1\over2}\ln{m_{H}^{2}\over 
	 p_{\perp}^{2}}\right)^{k+1}\ ,
	\label{eq:lnsqomz} 
\end{eqnarray}
where we recall that $\delta=p_\perp/(m_\perp+p_\perp)$.  For the second
case of inner integrals of the form $\int_x^{1}dz$ we require the following
identities as $p_\perp/m_H
\rightarrow0$:
\begin{eqnarray}
	{1\over1-z+\lambda} 
	 & \longrightarrow &
	 \left[{1\over1-z}\right]_{+}
	-\delta(1-z)\left[\phantom{1\over1}\!\!\!\!\ln\lambda\right]\ ,\nonumber\\
	{\ln(1-z+\lambda)\over(1-z)_{+}} 	 
& \longrightarrow & 
	\left[{\ln(1-z)\over1-z}\right]_{+}
	+\delta(1-z)
	\left[{1\over2}\ln^2\lambda+{\pi^{2}\over6}\right]\ ,\nonumber\\
	{\ln(1-z)\over1-z+\lambda} 	 
& \longrightarrow & 
	\left[{\ln(1-z)\over1-z}\right]_{+}
	-\delta(1-z)
	\left[{1\over2}\ln^2\lambda+{\pi^{2}\over6}\right]\ ,\nonumber\\
	{\ln(1-z+\lambda)\over1-z+\lambda} 	 
& \longrightarrow & 
	\left[{\ln(1-z)\over1-z}\right]_{+}
	-\delta(1-z)
	\left[{1\over2}\ln^2\lambda\right]\ ,\nonumber\\
	{\ln(1-z+\lambda)\over1-z+\tilde\lambda} 	 
& \longrightarrow & 
	\left[{\ln(1-z)\over1-z}\right]_{+}
	-\delta(1-z)
	\left[{1\over2}\ln^2\tilde\lambda+{\rm Li}_2\left(1-{\lambda\over\tilde\lambda}\right)\right]\ ,\nonumber\\
	{\ln(1-z+\lambda)-\ln\lambda\over1-z} 	 
& \longrightarrow & 
	\left[{\ln(1-z)\over1-z}\right]_{+}
-{\ln\lambda\ \over(1-z)_+}
	+\delta(1-z)
	\left[{1\over2}\ln^2\lambda+{\pi^{2}\over6}\right]\ ,
	\label{eq:omzp} 
\end{eqnarray}
where $\lambda$ and $\tilde\lambda$ are ${\cal O}(p_\perp^2/m_\perp^2)$.

The gluon-gluon contributions have only one additional complication
beyond these.  On integrating lines 5--6 and lines 9--10 of
eq.~(\ref{eq:ggsing}), one encounters double convolutions, which
can be transformed into single convolutions.  However, extra care
must be taken with the endpoints of the integrals as $p_\perp/m_H\rightarrow0$.
In the process we require the following identities in this limit:
\begin{eqnarray}
	\int_{x_{i}^{0}}^{1-\delta}dz^\prime_{i}\,F(z^\prime_{i})\,
{1\over1-z^\prime_i}\,
\ln^2\left({1-z^\prime_i-\delta\over1-z^\prime_i}\right)
	\quad & \longrightarrow &\quad	 2\zeta(3)F(1)\ ,\nonumber
\end{eqnarray}
\begin{eqnarray}
\lefteqn{	\int_{x_{i}^{0}}^{1-\delta}dz^\prime_{i}\,F(z^\prime_{i})\,
{1\over1-z^\prime_i}\,
\left[-{\rm Li}_2\left({1-z^\prime_i-\delta\over1-z^\prime_i}\right)\right]
	\quad\longrightarrow\qquad }\nonumber\\
&&\qquad\qquad\quad -{\pi^2\over6} 
	\int_{x_{i}^{0}}^{1}dz^\prime_{i}\,F(z^\prime_{i})\,
{1\over(1-z^\prime_i)_+}
	\, +\,\left[-{\pi^2\over12}\ln^2{m_H^2\over p_\perp^2}\,+2\zeta(3)\right]
\,F(1)\ .
\end{eqnarray}

We have verified that the coefficients $_mC_n^{(gq)}$ and $_mC_n^{(gg)}$,
derived from our NLO cross section agree with that of eq.~(\ref{eq:mCn}).
As an example, we display separately the ``singular'' and ``nonsingular''
contributions to $_2C_0^{(gg)}$.   The singular contribution gives 
\begin{eqnarray}
	_{2}C_{0}^{(s,gg)}& =& -\Biggl[\Bigl(P_{gg}\circ f_{g}\Bigr)({\bar x}_{a}^{0})
\Bigl(P_{gg}\circ f_{g}\Bigr)({\bar x}_{b}^{0})+\Bigl(P_{gg}\circ P_{gg}\circ f_{g}\Bigr)({\bar x}_{a}^{0})
f_{g}({\bar x}_{b}^{0})\Biggr]\ln{\mu_F^2\over m_H^2}\nonumber\\
&& +\Biggl[2\beta_0\Bigl(P_{gg}\circ f_{g}\Bigr)({\bar x}_{a}^{0})
f_{g}({\bar x}_{b}^{0})-(2n_f)\Bigl(P_{gq}\circ P_{qg}\circ f_{g}\Bigr)({\bar x}_{a}^{0})
f_{g}({\bar x}_{b}^{0})\Biggr]\ln{\mu_F^2\over m_H^2}\nonumber\\
&& 
+\left[\Delta+N_c\pi^2+3\beta_0\ln{\mu_R^2\over m_H^2}\right]
\Bigl(P_{gg}\circ f_{g}\Bigr)({\bar x}_{a}^{0})f_{g}({\bar x}_{b}^{0})\nonumber\\
&&
+\left[4N_c^2\zeta(3)-3\beta_0^2\ln{\mu_R^2\over m_H^2}
-{3\beta_0\over2}\left(\Delta+N_c\pi^2\right)\right]
f_{g}({\bar x}_{a}^{0})f_{q}({\bar x}_{b}^{0})
\nonumber\\
&&
+(2n_f)\Bigl(C_{gq}\circ P_{qg}\circ f_{g}\Bigr)({\bar x}_{a}^{0})f_{g}({\bar x}_{b}^{0})
+\Bigl(D^{(s)}_{gg}\circ f_{g}\Bigr)({\bar x}_{a}^{0})
f_{q}({\bar x}_{b}^{0})\nonumber\\
&&+\{a\leftrightarrow b\}
	\ ,
\end{eqnarray}
where
\begin{eqnarray}
D^{(s)}_{gg}(z)&=&
{p_{gg}(z)\over(1-z)_+}\Biggl[
N_c{\rm Li}_2(1-z)+	\left(
	{67\over18}N_c-{5\over9}n_f\right)
-{\pi^2\over6}N_c
+2\beta_0\ln{z}
\Biggr]\nonumber\\
&&+\beta_0\Biggl[\left({1\over2}\Delta+{2\pi^2\over3}N_c\right)\delta(1-z)-N_c{(1-z)^3\over z}\ln{z}
\Biggr]\nonumber\\
&&-{N_c\over3}(N_c-n_f)z\nonumber\\
&&
-\Bigl([P_{gg}\ln{x}]\circ P_{gg}\Bigr)(z)
-(2n_f)\Bigl(\left[P_{gq}\ln(1-x)\right]\circ P_{qg}\Bigr)(z)\nonumber\\
&&
+(2n_f)\Bigl(P_{gq}\circ \left[P_{qg}\ln(1-x)\right]\Bigr)(z)
-(2n_f)\Bigl(P_{gq}\circ \left[P_{qg}\ln{x}\right]\Bigr)(z)\nonumber\\
&&
+(2n_f)\Bigl(P_{gq}\circ C^{(1)}_{qg}\Bigr)(z)
-(2n_f)\Bigl(C^{(1)}_{gq}\circ P_{qg}\Bigr)(z)\nonumber\\
&=&{p_{gg}(z)\over(1-z)_+}\Biggl[N_c\ln^2{z}
-2N_c\ln{z}\ln(1-z)\nonumber\\
&&\qquad\qquad+N_c{\rm Li}_2(1-z)+	\left(
	{67\over18}N_c-{5\over9}n_f\right)
-{\pi^2\over6}N_c
+\beta_0\ln{z}
\Biggr]\nonumber\\
&&+\beta_0\Biggl[\left({1\over2}\Delta+{2\pi^2\over3}N_c\right)\delta(1-z)
-N_c{(1-z)^3\over z}\ln{z}
\Biggr]\nonumber\\
&&-{N_c\over3}(N_c-n_f)z\nonumber\\
&&+N_c^2\Biggl[\Biggl({22\over3}{(1-z)^3\over z}+16(1-z)\biggr)\ln{z}
+4(1+z)\ln^2{z}
\Biggr]\nonumber\\
&&
+n_fC_F\Biggl[
-(1+z)\ln^2{z}-(1+3z)\ln{z}+{2\over3z}-3-z+{10\over3}z^2
\Biggr]
\ .
\end{eqnarray}
The double convolutions have been transformed into single
convolutions in the second form for $D^{(s)}_{gg}(z)$.

We can write the ``nonsingular'' contribution as
\begin{eqnarray}
	_{2}C_{0}^{(ns,gg)}& =& 
\Bigl(D^{(ns)}_{gg}\circ f_{q}\Bigr)({\bar x}_{a}^{0})
f_{q}({\bar x}_{b}^{0})
	\ .\nonumber
\end{eqnarray}
The convolution function is a sum over
terms coming from each of the squared helicity amplitudes:
\begin{eqnarray}
D^{(ns)}_{gg}(z)&=&N_c^2\Biggl(
D_{A_0}(z)+D_{[A_{(1234)}+A_{(2341)}]}(z)+D_{[A_{(3412)}+A_{(4123)}]}(z)
\nonumber\\
&&\qquad
+D_{[A_{(1324)}+A_{(2413)}]}(z)
+D_{A_{(3241)}}(z)
+D_{A_{\epsilon}}(z)
\Biggr)\\
&&+n_fC_F\Biggl(D_{B_{1(+-)}}(z)+D_{B_{1(++)}}(z)\Biggr)
+n_fN_c\Biggl(D_{B_{2(+-)}}(z)+D_{B_{2(++)}}(z)\Biggr)\ ,\nonumber
\end{eqnarray}
where the labels 
correspond to the terms in eq.~(\ref{eq:ggterms}).
We obtain
\begin{eqnarray}
D_{A_0}(z)&=&{67\over18}{(1-z)^3\over z}\ ,\nonumber\\
D_{A_{(1234)}+A_{(2341)}}(z)&=&-{35\over9}{(1-z)^3\over z}
-{2\over3}(1-z)+\left[-{11\over6}{(1+z)^3\over z}+7(1+z)\right]\ln{z}
\nonumber\\
&&-{(1-z)^3\over z}{\rm Li}_2(1-z)-{1\over2}{(1-z)^3\over z}\ln^2{z}\ ,\nonumber\\
D_{A_{(3412)}+A_{(4123)}}(z)&=&-{1\over6}{(1-z)^3\over z}
+{19\over3}(1-z)+{1\over3}\nonumber\\
&&-\left[{1+z^4\over z(1-z)}\right]{\rm Li}_2(1-z)
-{1\over2}\left[{1+z^4\over z(1-z)}\right]\ln^2{z}\nonumber\\
&&+\left[-{11\over6}{(1+z)^3\over z}-{11\over6}
\left[{1+z^4\over z(1-z)}\right]+{17\over3}(1+z)\right]\ln{z}\ ,\nonumber\\
D_{A_{(1324)}+A_{(2413)}}(z)&=&-{35\over9}{(1-z)^3\over z}
-{8\over3}(1-z)-\left[{1+z^4\over z(1+z)}\right]S_2(z)\nonumber\\
&&+\left[-{11\over6}{(1+z)^3\over z}
+6(1+z)\right]\ln{z}\ ,\nonumber\\
D_{A_{(3241)}}(z)&=&-{35\over9}{(1-z)^3\over z}
-{73\over6}(1-z)-\left[{(1+z)^3\over z}\right]S_2(z)\nonumber\\
&&+\left[-{11\over6}{(1+z)^3\over z}
+(1+z)\right]\ln{z}\ ,\nonumber\\
D_{A_\epsilon}(z)&=&{2\over3}{(1-z)^3\over z}\ ,\nonumber\\
D_{B_{1(+-)}}(z)&=&0\ ,\nonumber\\
D_{B_{1(++)}}(z)&=&-5(1-z)-2(1+z)\ln{z}\ ,\nonumber\\
D_{B_{2(+-)}}(z)&=&-{13\over18}{(1-z)^3\over z}\ ,\\
D_{B_{2(++)}}(z)&=&-{13\over18}{(1-z)^3\over z}-3(1-z)-{1\over3}
+\left[{1\over3}{1+z^4\over z(1-z)}-{2\over3}(1+z)\right]\ln{z}\ ,\nonumber
\end{eqnarray}
where
\begin{equation}
S_2(z)\ =\ \int_{z\over1+z}^{1\over1+z}{dx\over x}\ln\left({1-x\over x}\right)
\ =\ {1\over2}\ln^2{z}+\int_z^1{dx\over x}\ln(1-x+z)\ .
\end{equation}
Note that the ${\cal O}(\epsilon)$ term $A_\epsilon$ contributes and is
crucial to obtaining the correct
small-$p_\perp$ limit.

Combining $_2C_0^{(s,gg)}+\,_2C_0^{(ns,gg)}$ and comparing with
the coefficient in eq~(\ref{eq:mCn}) obtained from the resummation, we
recognize the combination
\begin{eqnarray}
{1\over2}B^{(2)}\delta(1-z)+P^{(2)}_{gg}(z)&=&D^{(s)}_{gg}(z)+D^{(ns)}_{gg}(z)\nonumber\\
&=&{p_{gg}(z)\over(1-z)_+}
\Biggl[{1\over2}N_c\ln^2{z}
-2N_c\ln{z}\ln(1-z)+{67\over18}N_c-{5\over9}n_f-{\pi^2\over6}N_c\Biggr]\nonumber\\
&&+N_c^2\Biggl[\Biggl(-{25\over3}+{11\over3}z-{44\over3}z^2\biggr)\ln{z}
+4(1+z)\ln^2{z}\nonumber\\
&&\qquad\quad-\left({1+z^4+(1+z)^4\over z(1+z)}\right)S_2(z)
+{67\over9}\left(z^2-{1\over z}\right)+{27\over2}(1-z)
\Biggr]\nonumber\\
&&+n_fN_c\Biggl[-{2\over3}(1+z)\ln{z}
+{13\over9}\left(z^2-{1\over z}\right)+1-z
\Biggr]\nonumber\\
&&
+n_fC_F\Biggl[
-(1+z)\ln^2{z}-(3+5z)\ln{z}+{2\over3z}-8+4z+{10\over3}z^2
\Biggr]\nonumber\\
&&+\beta_0\delta(1-z)\Biggl[{1\over2}\left(\Delta+{4\pi^2\over3}N_c\right)
\Biggr]
\ .
\end{eqnarray}
This agrees with the result for $B^{(2)}$ from Ref.~\cite{dFG} and
the result for $P^{(2)}_{gg}(z)$, given in Ref.~\cite{ESW}.


\end{document}